\documentclass[aps,pra,amsmath,amssymb,reprint,groupedaddress]{revtex4-2}
\usepackage[utf8]{inputenc}
\usepackage[english]{babel} 
\usepackage[T1]{fontenc}
\usepackage{graphicx}
\usepackage{placeins}
\usepackage{amsmath,amssymb}
\usepackage[colorlinks=true,citecolor=red]{hyperref}		
\usepackage{natbib}
\usepackage{xcolor}
\usepackage{siunitx}
\usepackage{braket}
\newcommand\OD{\mathrm{OD}}
\renewcommand\vr{\mathbf{r}}
\newcommand\NEP{P_{\rm NEP}}
\newcommand\SNR{\mathrm{SNR}}
\newcommand\coh{\rho_{eg}}
\newcommand\Rb{$^{87}$Rb}

\begin{document}

\title{Optimal strategies for low-noise detection of atoms using resonant frequency modulation spectroscopy in cold atom interferometers}
\author{Ryan J. Thomas}\email{ryan.thomas@anu.edu.au}
\author{Samuel Legge}
\author{Simon A. Haine}
\author{John D. Close}
\affiliation{Department of Quantum Science and Technology, Research School of Physics, Australian National University, Canberra 2601, Australia}

\begin{abstract}
Resonant frequency modulation spectroscopy has been previously used as a highly-sensitive method for measuring the output of cold atom interferometers.  Using a detailed model that accounts for optical saturation, laser intensities and atomic densities that vary spatially, and radiation pressure on the atoms, we theoretically investigate under what parameter regimes the optimum signal-to-noise ratio is found under experimentally realistic conditions.  We compare this technique to the standard method of fluorescence imaging and find that it outperforms fluorescence imaging for compact interferometers using condensed atomic sources or where the photon collection efficiency is limited.  However, we find that fluorescence imaging is likely to be the preferred method when using squeezed atomic sources due to limited atom number.

\end{abstract}

\maketitle

\section{Introduction}

Cold atom interferometers (CAIs) perform measurements by encoding a quantity of interest as the phase difference between different paths of the interferometer, and they have been used to test the weak equivalence principle \cite{asenbaum_atom-interferometric_2020}, to measure fundamental constants such as the gravitational coupling constant $G$ \cite{fixler_atom_2007,rosi_precision_2014}, the fine structure constant $\alpha$ \cite{weiss_precision_1994,lan_clock_2013,yu_atom-interferometry_2019}, and the tune-out wavelength of $^{87}$Rb \cite{fallon_measurement_2022,leonard_high-precision_2015}, and to make absolute measurements of gravity \cite{freier_mobile_2016}, gravity gradients \cite{mcguirk_sensitive_2002,stray_quantum_2022}, and rotation \cite{canuel_six-axis_2006,berg_composite-light-pulse_2015}.  Regardless of the configuration or quantity to be measured, the interferometer phase is inferred by counting the relative number of atoms in each output port of the interferometer.  Uncertainty in counting atoms translates directly to uncertainty in the phase and thus in the quantity of interest.  Therefore, ensuring that detection noise is a negligible contribution to the overall phase uncertainty is a critical part of designing CAIs.  Reducing the detection noise becomes especially important for quantum-enhanced CAIs using entangled atom sources which have detection noise requirements below the atomic shot noise limit \cite{szigeti_high-precision_2020,corgier_delta-kick_2021,kritsotakis_spin_2021}.  

\begin{figure}[b]
	\centering
	\includegraphics[width=\columnwidth]{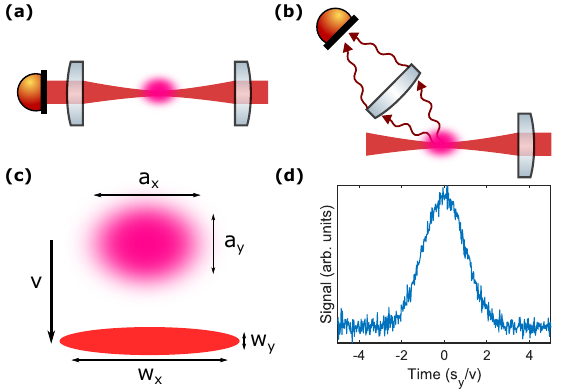}
	\caption{\textbf{(a-b)} Schematic of a time-of-flight detection setup.  A sample of atoms (purple) falls through a Gaussian laser beam that has been focused using a cylindrical lens.  \textbf{(a)} Absorption imaging and FMI measures the optical power in the incident beam.  \textbf{(b)} FI collects scattered light from the sample and focuses it using a single lens onto a photodetector.  \textbf{(c)} Geometry of the sample and laser beam.  The sample falls in the $y$ direction with velocity $v$.  \textbf{(d)} Simulated FMI signal for a sample of $10^5$ atoms with a temperature of \SI{100}{\nano\kelvin} at a time of flight of \SI{200}{\milli\second}.}
	\label{fg:schematic}
\end{figure}

We consider here optical detection of atoms using a time-of-flight measurement \cite{hardman_time--flight_2016,cheng_momentum-resolved_2018,gauguet_characterization_2009} where the atoms fall through a near-resonant laser beam as shown in Fig.~\ref{fg:schematic}.  One either measures the absorption of light (Fig.~\ref{fg:schematic}a) as it passes through the atoms or the fluorescence of the atoms (Fig.~\ref{fg:schematic}b), and the difference between the two measurements lies in their scaling with optical depth ($\OD$).  As a simple example, suppose we have a uniform sample of atoms with an on-resonance $\OD \ll 1$.  If we illuminate this sample with a resonant optical field of power $P_i$, then the amount of power scattered from the field is $P_{\rm sc} = [1 - e^{-\OD}] P_i \approx \OD P_i$.  The signal of interest in both absorption imaging (AI) and fluorescence imaging (FI) arises from the same scattered field, but in FI the variance in the detected power is proportional to the scattered power $P_{\rm sc}$ while for AI the variance is proportional to the incident power $P_i$.  If we assume that fluorescence photons are collected with a solid angle efficiency $\eta$, where typically $\eta < 0.1$, then the photon shot noise limited signal-to-noise ratios (SNRs) for FI and AI are
\begin{subequations}\label{eq:rough-snrs}
\begin{align}
	\mathrm{SNR}_{\rm FI} &\propto \sqrt{\eta \OD P_i}\label{eq:rough-fluorescence-snr}\\
	\mathrm{SNR}_{\rm AI} &\propto \OD\sqrt{P_i}\label{eq:rough-absorption-snr}
\end{align}
\end{subequations}
which imply that for $\eta > \OD$ FI has superior SNR compared to AI.  This condition is readily realised in both laboratory-based and mobile CAIs which use thermal atom sources with low $\OD$ and has led to FI becoming the standard read-out method for CAIs \cite{freier_mobile_2016,wu_gravity_2019,stray_quantum_2022}.  However, certain use-cases for CAIs, such as their application in highly dynamic environments, may benefit from short interrogation times and/or Bose-condensed atomic sources \cite{rudolph_high-flux_2015} where the condition $\OD > \eta$ is satisfied, and therefore AI may have superior performance to FI. Notably, Bose-condensed samples subject to delta-kick collimation can retain $\OD > 1$ for long expansion times \cite{heine_transportable_2020,deppner_collective-mode_2021}, and appropriately AI is used in these experiments.

Simply measuring the absorption of a single laser frequency as it passes through a sample of atoms is a poor measurement strategy, as the change in the optical power due to the atoms is typically small compared to the input optical power, yielding a mismatch between the dynamic range of the signal and the detector.  An alternative method was proposed \cite{hardman_time--flight_2016} which used resonant frequency modulation spectroscopy \cite{bjorklund_frequency_1983} to generate a radio-frequency (RF) signal which could be subsequently amplified to match the dynamic range of the measurement device.  This technique, which the authors called frequency modulation imaging (FMI), involves phase-modulating a laser beam and setting the laser frequency such that one of the sidebands is resonant with the atomic transition, so that absorption of the resonant sideband converts the phase modulation into amplitude modulation.  An AC-coupled detector then ideally only measures a signal at the modulation frequency when the atoms are present.  SNRs of nearly $10^3$ were measured with $2\times 10^6$ Bose-condensed atoms after \SI{750}{\milli\second} of expansion \cite{hardman_time--flight_2016}.  Subsequently, FMI has also found application in detecting laser-cooled atomic sources \cite{saywell_enhancing_2023}.

In this paper, we theoretically investigate how to obtain the maximum SNR for FMI and in what scenarios FMI is expected to have a higher SNR than FI.  Optical saturation, spatial variations in the laser beam and atomic density, and radiation pressure effects are included in our model.  We show that compact CAIs with short expansion times and with realistic atom numbers and temperatures are likely to benefit from using FMI as the detection method.

\section{Theoretical model}
\label{sec:theory}

\subsection{Fluorescence and absorption imaging}

We consider here a two-level atom with ground state $\ket{g}$ and excited state $\ket{e}$ separated by an electric dipole-allowed transition with dipole moment $d$ and frequency $\omega_0$.  Although the atoms used in CAIs are multi-level atoms, the most commonly used atoms are the alkali metal atoms $^{87}$Rb and $^{133}$Cs which have large excited-state splittings and nearly-closed transitions between the ground and excited states corresponding to the $\ket{S_{1/2},F,m_F = \pm F}\to\ket{P_{3/2},F' = F + 1,m_F' = m_F \pm 1}$ transitions.  Incident on the atoms is an electric field $\mathcal{E}(\vr,t) = E(\vr)e^{-i\omega t} + \textrm{c.c.}$ with frequency $\omega$ and spatially-varying amplitude $E(\vr)$.  We assume that the quantization axis of the atoms is aligned with the propagation direction of the laser beam and that the electric field is $\sigma^+$ polarized.  Although atoms in CAIs typically start in magnetically-insensitive ground state sublevels, they are quickly optically pumped into the highest angular momentum state \footnote{For $^{87}$Rb atoms starting in the $\ket{F=2,m_F=0}$ state, the timescale for optical pumping into the $\ket{2,2}$ state is $\mathord{\sim}\SI{10}{\micro\second}$ for a resonant laser whose intensity is one-tenth the saturation intensity.} where there exists only a single allowed electric dipole transition for $\sigma^+$ polarized light.  We also assume that external magnetic and electric fields are weak, so that mixing of the excited states is negligible.  Thus, in the rotating-wave approximation, we can write the Hamiltonian as
\begin{equation}
    \hat{H} = \hbar\begin{bmatrix}
        0 & -\Omega^*(\vr)\\
        -\Omega(\vr) & -\Delta
    \end{bmatrix}
    \label{eq:hamiltonian}
\end{equation}
with detuning $\Delta$ and Rabi frequency $\Omega(\vr) = dE(\vr)/\hbar$.  Noting that the total population $\rho_{gg} + \rho_{ee} = 1$, where $\rho_{ij}$ is the $ij$'th component of the density matrix $\hat{\rho}$, we can write the time evolution of the density matrix in terms of the population imbalance $w = \rho_{ee} - \rho_{gg}$ and the coherence $\coh$ as \cite{laser_cooling_and_trapping}
\begin{subequations}\label{eq:time-evolution}
	\begin{align}
	    \dot{\rho}_{eg}(t) &= i\left(\Delta + i\frac{\Gamma}{2}\right)\coh(t) - i\Omega w(t)\label{eq:coherence-evolution}\\
	    \dot{w}(t) &= -\Gamma[1 + w(t)] + 2i\Omega \coh^*(t) - 2i\Omega^*\coh(t)\label{eq:imbalance-evolution}
	\end{align}
\end{subequations}
where $\Gamma$ is the full-width at half-maximum of the transition, and we have dropped the spatial dependence for brevity.  For atom-light interaction times much longer than $\Gamma^{-1} \approx \SI{25}{\nano\second}$, we can use the steady-state solutions to Eqs.~\eqref{eq:coherence-evolution} and \eqref{eq:imbalance-evolution} to get the coherence and excited state population
\begin{subequations}
	\begin{align}
	    \coh(\vr) &= \Omega(\vr)\frac{i\Gamma/2 - \Delta}{\Gamma^2/4 + \Delta^2 + 2|\Omega(\vr)|^2}\label{eq:coherence-steady-state}\\
	    \rho_{ee}(\vr) &= \frac{|\Omega(\vr)|^2}{\Gamma^2/4 + \Delta^2 + 2|\Omega(\vr)|^2}.\label{eq:excited-steady-state}
	\end{align}
\end{subequations}

Radiation pressure from the incident laser provides a net acceleration of the atoms in the direction of the laser, and this change in the atoms' velocities causes a spatio-temporal change in their detunings.  Each absorbed photon changes an atom's detuning by $\omega_{\rm rec} = \hbar k^2/m$ on average, with $k$ the wavenumber of the laser.  In the steady-state the rate at which an atom scatters/absorbs photons is $\Gamma\rho_{ee}(\vr)$; thus, the detuning changes as $\dot{\Delta}(\vr,t) = -\omega_{\rm rec}\Gamma\rho_{ee}(\vr,t)$, which has the implicit solution
\begin{equation}
	\Delta(\vr,t) + \frac{4\Delta^3(\vr,t)}{3(1 + s(\vr))\Gamma^2} = -\frac{\omega_{\rm rec}\Gamma t}{2}\frac{s(\vr)}{1 + s(\vr)}
	\label{eq:change-in-detuning}
\end{equation}
for saturation parameter $s = 8|\Omega|^2/\Gamma^2 = I/I_{\rm sat}$ for incident intensity $I$ and saturation intensity $I_{\rm sat} = \Gamma\hbar c k^3/(12\pi)$.  As the timescale for changes in detuning to occur are $\omega_{\rm rec}^{-1} \gg \Gamma^{-1}$, we are justified in using the steady-state solutions for $\coh$ and $w$ to compute the change in detuning.  The cubic term can be neglected for both $\omega_{\rm rec}t \ll 1$ and $s \gg 1$, as these correspond to situations where there is insufficient acceleration to change the detuning by more than either the natural or power-broadened linewidth.

For the specific application of terrestrial CAIs, we assume that the atomic sample is moving in the $y$ direction with velocity $v$ and has an optical depth $\OD \ll 1$ that weakly perturbs the incident light.  In FI, we detect the photons scattered from the atoms where the local scattering rate is $\Gamma\rho_{ee}(x,y,t)$, and the temporal variation is due to the change in detuning.  Assuming we collect scattered photons with solid angle efficiency $\eta$, the measured scattered power as a function of time is
\begin{equation}
	P_{\rm sc}(t) = 2\eta I_{\rm sat} \int \OD(x,y - vt)\rho_{ee}(x,y,t) dxdy.
	\label{eq:fluorescence-scattered-power}
\end{equation}

In AI, we measure the total intensity of the field in the direction of the original beam, $I(x,y,t) = 2\epsilon_0 c |E(x,y,t)|^2$, after the laser has passed through the atoms.  For $\OD \ll 1$, the electric field can be written as the sum of the incident field $E_0(x,y)$ and the change in the field $\Delta E(x,y,t)$ due to the atoms where
\begin{equation}
	\Delta E(x,y,t) = \frac{i\hbar\Gamma}{4d}\OD(x,y - vt)\coh(x,y,t)
 	\label{eq:electric-field-change}
\end{equation}
in the paraxial limit and where we neglect diffraction of the light as it passes through the atomic sample.  The total measured power is then
\begin{equation}
	P(t) = P_i + 4\epsilon_0 c\mathrm{Re}\left(\int E_0^*(x,y)\Delta E(x,y,t) dxdy\right)
	\label{eq:ai-power}
\end{equation}
for incident power $P_i$, and where we have neglected the quadratic term $\Delta E^2(x,y,t)$ since $\Delta E \ll E_0$ when $\OD \ll 1$.  The change in the measured power occurs at baseband, which makes it more susceptible to $1/f$ noise and means it cannot be easily separated from $P_i$ for amplification to match the detector's dynamic range.

\subsection{Frequency modulation imaging}
\label{ssec:fmi-theory}

FMI makes use of a phase or frequency modulated field so that $E(\vr,t) = \sum_{n = -\infty}^\infty E_n(\vr)e^{in\omega_m t}$ for modulation frequency $\omega_m$ and sideband amplitudes $E_n(\vr)$ which are determined by the Jacobi-Anger expansion.  In FMI, we choose the first order sideband $E_1$ at $\omega - \omega_m$ to be on-resonance and $\omega_m$ to be large enough that $E_1$ is the only frequency component of the field to interact with the atoms; we will relax this assumption in Section \ref{ssec:constrained-model}.  We also assume that the carrier is blue-detuned from the transition, so that for typical CAIs using $^{87}$Rb or $^{133}$Cs the strong carrier $E_0$ is not resonant with other excited state manifolds.  FMI in this situation is exactly absorption imaging but on the resonant sideband.  The signal of interest is then the power at $\omega_m$ defined by $P_{\omega_m}(t) = \int I_{\omega_m}(x,y,t) dxdy$ where
\begin{equation}
	I_{\omega_m}(x,y,t) = 4\epsilon_0 c\big|E_0^*\Delta E_1\big|\cos(\omega_m t + \phi)
	\label{eq:intensity-ac}
\end{equation}
with $\phi$ the phase of $E_0^*\Delta E_1$ and $\Delta E_1$ the change in the $-\omega_m$ field due to the atoms.  Detection of this signal is done by demodulating at $\omega_m$ with a free choice of the demodulation phase, which means that the maximum intensity signal at $\omega_m$ is simply the amplitude of Eq.~\eqref{eq:intensity-ac}.  We note that realistic phase modulators inevitably have some level of residual amplitude modulation (RAM), which will lead to a signal at $\omega_m$ in the absence of atoms; however, RAM is small in modern modulators, and it only adds a constant offset to the measured signal at $\omega_m$ which can be measured when there are no atoms.  Therefore, we will neglect the effects of RAM in the remainder of this article.

The optical power for both FMI and FI is measured using a photodetector, and we model our photodetector as an avalanche photodetector (APD) with responsivity $R_0$, noise-equivalent power (NEP) $\NEP$, electron multiplication factor $M$, and hole-to-electron ionization ratio $\kappa$.  The two-sided, real-frequency power spectral density for the current generated by the APD for a given total power $P$ is
\begin{equation}
    S_{ii} = (R_0 M \NEP)^2 + e F M^2 R_0 P
    \label{eq:psd-apd}
\end{equation}
with excess noise factor $F = \kappa M + (1-\kappa)(2 - M^{-1})$ \cite{mcintyre_multiplication_1966}.  Note that when $M = 1$, $F = 1$ regardless of $\kappa$, which means that Eq.~\eqref{eq:psd-apd} can also model the noise generated by non-multiplying photodiodes.  Assuming that we have a detection bandwidth $B$ associated with an effective averaging time $\tau = (2B)^{-1}$, the peak SNRs for FMI and FI are
\begin{subequations}\label{eq:snrs}
\begin{align}
    \SNR_{\rm FMI} &= \frac{P_1\sqrt{\tau}}{\sqrt{\NEP^2 + eFP/R_0}}\label{eq:fmi-snr}\\
    \SNR_{\rm FI} &= \frac{P_{\rm sc} \sqrt{\tau}}{\sqrt{\NEP^2 + eFP_{\rm sc}/R_0}}\label{eq:fi-snr}
\end{align}
\end{subequations}
where $P$ is the total power incident on the detector.  The SNR for counting the total number of atoms can be obtained from Eq.~\eqref{eq:snrs} by replacing $\tau$ with the effective temporal width of the time of flight signal, $\tau_s$.

\section{Results}
\label{sec:results}

\subsection{Uniform optical depth and no radiation pressure}
\label{ssec:unconstrained-uniform-density}

We first consider the simple scenario where the optical depth is uniform and the incident laser has a Gaussian intensity profile with horizontal and vertical $e^{-2}$ radii $w_x$ and $w_y$ centered at the origin (Fig.~\ref{fg:schematic}c).  We also assume that there is no change in the detuning due to radiation pressure.  The optimal SNRs for both FMI and FI occur when $\Delta = 0$ corresponding to maximum absorption/scattering of light.  The intensity at $\omega_m$ for FMI is then
\begin{equation}
	I_{\omega_m}(x,y) = \OD I_{\rm sat}\sqrt{s(x,y)} \frac{\sqrt{s_1(x,y)}}{1 + s_1(x,y)}
	\label{eq:simple-first-order-fmi-intensity}
\end{equation}
where $s_1(x,y) = I_1(x,y)/I_{\rm sat}$ is the saturation parameter associated with the intensity in the on-resonance sideband at $-\omega_m$, and $s(x,y)$ is the saturation parameter associated with the total amount of power incident on the detector.  For uniform $\OD$, the signal power is computed by integrating Eq.~\eqref{eq:simple-first-order-fmi-intensity} over all space, noting that the spatial profile for both the carrier and sideband are the same, to yield the detected power as a function of $s_1(0,0)$
\begin{equation}
	P_{\omega_m}(s_1(0,0)) = \OD P_{\rm sat} \sqrt{s(0,0)} \frac{\log[1 + s_1(0,0)]}{\sqrt{s_1(0,0)}}
	\label{eq:simple-first-order-fmi-power}
\end{equation}
where $P_{\rm sat} = I_{\rm sat}\pi w_x w_y/2$ is the ``saturation power'' corresponding to the power required to have a peak saturation parameter of $1$. Equation \eqref{eq:simple-first-order-fmi-power} has a maximum when $s_1(0,0) \approx 3.92$, yielding a maximum signal power
\begin{equation}
	P_{\omega_m} \approx 0.8\times \OD P_{\rm sat}\sqrt{s(0,0)}.
	\label{eq:simple-first-order-fmi-signal}
\end{equation}
From Eqs.~\eqref{eq:excited-steady-state} and \eqref{eq:fluorescence-scattered-power} we see that the maximum scattered power for FI occurs when $s \rightarrow \infty$, yielding $P_{\rm sc} = \eta P_{\rm sat} \OD$.  In the limit where the detected optical power on the photodetector is large enough that the noise is dominated by photon shot noise, the ratio of SNRs is
\begin{equation}
	\frac{\SNR_{\rm FMI}}{\SNR_{\rm FI}} \approx \sqrt{\frac{\OD}{1.56\times\eta}},
	\label{eq:simple-ratio-optimum-snrs}  
\end{equation}
and we can clearly see that FMI is nominally superior to FI when $\OD > 1.56\times\eta$ when using the same photodetector.  A typical range for $\eta$ in CAIs is $0.01 <\eta < 0.1$, with the high end of the range often requiring the use of two windows and two photodetectors \cite{li_continuous_2023}.  As an example, in our apparatus we perform atom interferometry with BECs of $\sim\!10^6$ atoms with total expansion times ranging from \SI{200}{\milli\second} to \SI{730}{\milli\second}, corresponding to optical depths ranging from $0.45$ to $0.034$ \cite{hardman_simultaneous_2016}.  The maximum collection efficiency that we can achieve with a single vacuum window is $\eta \approx 0.05$, which means that we expect FMI to have higher $\SNR$ than FI for expansion times less than \SI{480}{\milli\second} and lower $\SNR$ for longer expansion times.  In contrast, many CAIs use laser-cooled, velocity-selected atomic sources for which $\OD < 0.1$ at modest expansion times, implying that for these sources FI likely has a higher SNR than FMI for similar collection efficiencies.

\subsection{Spatial effects and radiation pressure}
\label{ssec:unconstrained-spatial-dependence}

\begin{figure}[t]
	\centering
	\includegraphics[width=\columnwidth]{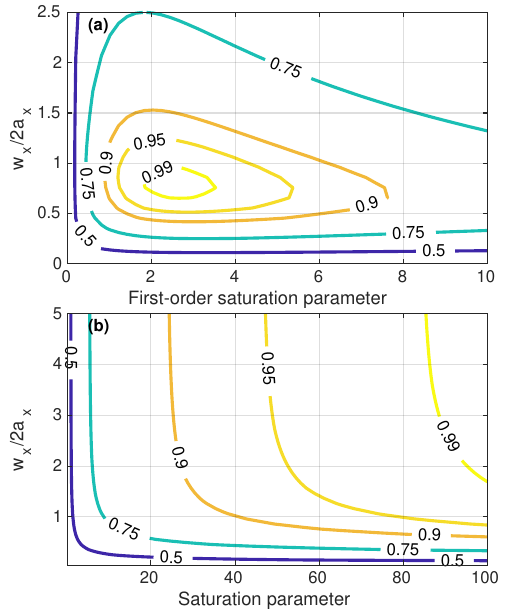}
	\caption{\textbf{(a)} Contour plot of the normalised $\SNR_{\rm FMI}$ for different first-order saturation parameters and horizontal beam waists normalised as $w_x/2a_x$.  The vertical beam waist has been fixed to be $w_y = 0.1a_y$.  \textbf{(b)} Contour plot of the normalised $\SNR_{\rm FI}$ for different saturation parameters and horizontal beam waists.  Both panels assume $N = 10^6$ \Rb{} atoms with a temperature of \SI{50}{\nano\kelvin}, $t_{\rm tof} = \SI{200}{\milli\second}$, and a \SI{20}{\kilo\hertz} detection bandwidth.  The SNR for FMI is calculated assuming a normal photodiode with $R_0 = \SI{0.46}{\ampere/\watt}$ with shot-noise limited detection, while the SNR for FI is calculated assuming a collection efficiency of $\eta = 0.05$ and an APD with $R_0 = \SI{0.53}{\ampere/\watt}$, excess noise factor $F = 2.4$, and $\NEP = 0$.  The maximum (peak) SNRs under these conditions are $2000$ for FMI and $700$ for FI.}
	\label{fg:simple-spatial-figure}
\end{figure}

\begin{figure*}[t]
	\centering
	\includegraphics[width=\textwidth]{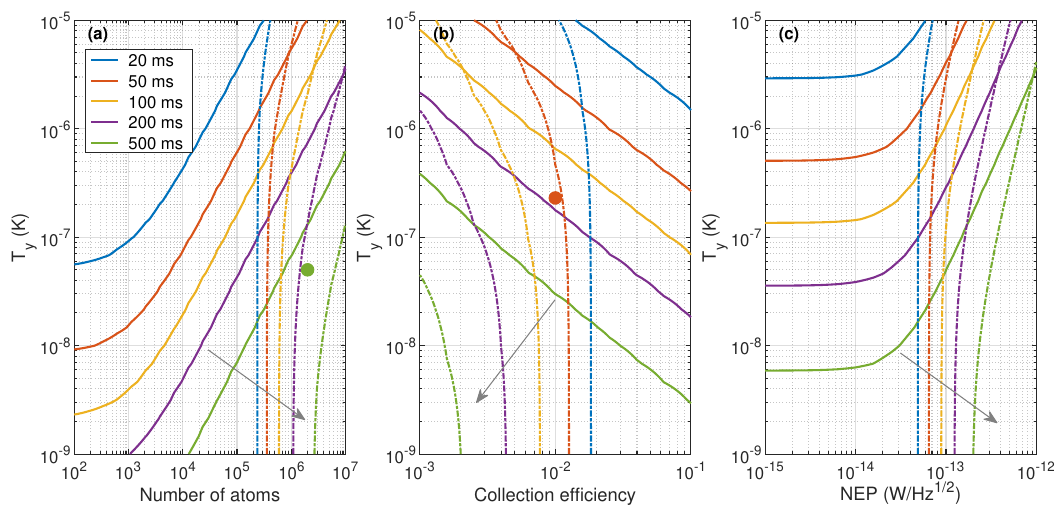}
	\caption{Contours where $\SNR_{\rm FMI} = \SNR_{\rm FI}$ for different $t_{\rm tof}$ as indicated in the legend and under different conditions for the two cases of no initial size of the sample and $T_x = T_y$ (solid lines), and initial sizes $a_x = a_y = \SI{500}{\micro\meter}$ and $T_x = \SI{3}{\micro\kelvin}$ (dashed-dotted lines).  We have assumed the atoms are \Rb{}.  \textbf{(a)} Contours of equal $\SNR$ for $\eta = 0.05$ and optimised choice of photodetectors (see text).  The green circle is indicative of parameters from Ref.~\cite{hardman_time--flight_2016} where FMI was used for detection of a BEC at an expansion time of \SI{530}{\milli\second}.  \textbf{(b)} Contours of equal $\SNR$ for $N = 10^5$ atoms and optimised choice of photodetectors.  The red circle is indicative of parameters from Ref.~\cite{saywell_enhancing_2023} where FMI was used for detection of a laser-cooled source at an expansion time of $\mathord{\sim}\SI{50}{\milli\second}$.  \textbf{(c)} Contours of equal $\SNR$ for $N = 10^5$ atoms and $\eta = 0.05$. Grey arrows indicate the direction where $\SNR_{\rm FMI} > \SNR_{\rm FI}$.}
	\label{fg:unconstrained-comparison}
\end{figure*}

Atomic sources for CAIs have spatially-dependent densities which are associated with the momentum width of the source after expansion by a time-of-flight $t_{\rm tof}$.  For sufficiently large $t_{\rm tof}$, the size of the atomic cloud $a_{x,y}$ (Fig.~\ref{fg:schematic}c) increases linearly with the time-of-flight as $a_{x,y} = \Delta v_{x,y} t_{\rm tof}$ where $\Delta v_{x,y} = \sqrt{k_B T_{x,y}/m}$ is the velocity width of the source in either the $x$ or $y$ directions for temperatures $T_{x,y}$ and mass $m$.  Our aim is to accurately reconstruct the 1D density distribution of the atoms after the interferometer, either because we wish to minimise spatial overlap of the two output ports to ensure accurate counting of atoms, or because the vertical spatial distribution contains important information such as when the interferometer phase is encoded using spatial fringes \cite{wigley_readout-delay-free_2019,wang_enhanced_2023,ben-aicha_dual_2024}.  Therefore, we fix our vertical beam waist $w_y$ to be a fraction $f_y$ of the vertical size of the cloud $a_y$, $w_y = f_ya_y$.  Since the maximum $\SNR$ for both FMI and FI is proportional to $w_y$, larger $f_y$ corresponds to higher SNR for both imaging types but with poorer vertical resolution.

The choice of vertical beam waist also impacts radiation pressure effects.  The atom-light interaction time is approximately $\tau_{\rm int} = w_y/v$, so that as the beam waist increases the atoms' detunings change more and a higher saturation parameter is required to ensure that the atoms' detunings remain within the power-broadened linewidth.  This effect is most pronounced for FMI, as the optimal saturation parameter is of order unity for negligible changes to detunings, while for FI the optimal saturation parameter is infinite.

The horizontal beam waist $w_x$ also affects the SNR for both FMI and FI, as the signal powers increase when more atoms are illuminated by the incident beam.  For FMI, however, when $w_x \gg a_x$ the edges of the beam do not interact with any atoms, so those sections of the beam contribute only noise to the measurement; therefore, there must be an optimal choice of $f_x = w_x/a_x$.  For FI, we want $w_x \gg a_x$ in order to illuminate as many atoms as possible since the parts of the incident beam that do not interact with the atoms contribute neither signal nor noise to the measured scattered power.

We assume that we have a sample of $N$ \Rb{} atoms, corresponding to a peak on-resonance optical depth $\OD_0$, and horizontal and vertical  $e^{-1/2}$ radii $a_x$ and $a_y$ (Fig.~\ref{fg:schematic}c).  We fix the vertical beam waist to be $w_y = 0.1 a_y$, which increases the measured source width by approximately $1\%$.  We numerically integrate Eqs.~\eqref{eq:fluorescence-scattered-power} and \eqref{eq:intensity-ac} over the spatial profiles of the atomic sample and the laser beam as the atomic sample moves through the beam, including the spatio-temporal effect of radiation pressure on the detuning of the atoms.  Under these conditions, we calculate $\SNR_{\rm FMI}$ from the maximum of the time-dependent signal as a function of horizontal beam waist and saturation parameter, and the results are shown in Fig.~\ref{fg:simple-spatial-figure}a.  The optimum parameters are $w_x \approx 1.5a_x$ and $s_1 \approx 2.5$, but near-optimal parameters can be found over a range of values.  As expected, the optimal saturation parameter is lower than in the uniform density case since there are fewer atoms in the wings of the beam which can contribute signal.

We additionally compute $\SNR_{\rm FI}$ under similar conditions in Fig.~\ref{fg:simple-spatial-figure}b.  In contrast to FMI, larger horizontal beam sizes are always advantageous provided sufficient power is available to ensure that the peak saturation parameter is sufficiently large.  A further difference is that the SNR monotonically increases with increasing peak saturation parameter, which is partially due to the $s/(1 + s)$ dependence of the scattered power and partially a result of the intense laser beam effectively having a larger size and thus covering more atoms.  As the beam power increases for a fixed size the temporal width of the measured signal will increase and can result in the signals from the two momentum modes overlapping.  We therefore fix the optimal peak saturation parameter for FI to be $100$, which results in a negligible $3\%$ increase in the measured vertical width relative to the actual width of the atomic sample.

Using the optimal parameters from Fig.~\ref{fg:simple-spatial-figure}, we calculate contours where $\SNR_{\rm FMI} = \SNR_{\rm FI}$ for two common CAI sources in Fig.~\ref{fg:unconstrained-comparison}: those with zero spatial size when dropped and with equal temperatures in both the $x$ and $y$ directions, such as condensed samples, and those with spatial sizes $a_x = a_y = \SI{500}{\micro\meter}$ when dropped and with a fixed temperature $T_x = \SI{3}{\micro\kelvin}$, such as laser-cooled samples subjected to velocity selection.  We consider variations in the vertical temperature $T_y$, the number of atoms, the FI collection efficiency, and the FI detector NEP.  In Fig.~\ref{fg:unconstrained-comparison}a, we show how the contours of equal $\SNR$ change with $T_y$ and $N$ for $\eta = 0.05$ and for an optimised choice of inexpensive and commercially available photodetectors: here, the Thorlabs DET02A for FMI with $\NEP = \SI{140}{\femto\watt/\hertz^{1/2}}$ and $R_{0,\rm FMI} = \SI{0.46}{\ampere/\watt}$, and the Thorlabs APD440A for FI with $\NEP = \SI{3.5}{\femto\watt/\hertz^{1/2}}$, $R_{0,\rm FI} = \SI{0.53}{\ampere/\watt}$, and $F = 2.4$ at $M = 10$.  For sources of equal $x$ and $y$ temperatures and zero initial size, there are broad regions of $T_y$ and $N$ where FMI has superior $\SNR$ compared to FI.  The asymptotic behaviors, corresponding to straight lines in Fig.~\ref{fg:unconstrained-comparison}a, are approximately determined by
\begin{equation}
	\frac{\SNR_{\rm FMI}}{\SNR_{\rm FI}} \approx \sqrt{\frac{\OD F R_{0,\rm FMI}}{2\eta R_{0,\rm FI}}}
	\label{eq:determined-ratio-optimum-snrs}  
\end{equation}
which is a generalisation of Eq.~\eqref{eq:simple-ratio-optimum-snrs} for non-identical photodetectors and including the spatial variation of the atomic density.  For laser-cooled samples we see that the contours of equal $\SNR$ are mostly temperature-independent as the size of the sample, and therefore the $\OD$, is dominated by the initial size of the sample.  Note that for any given time-of-flight a critical number of atoms must be present in the laser-cooled sample for $\SNR_{\rm FMI} > \SNR_{\rm FI}$ over a wide range of vertical temperatures.  We show in Fig.~\ref{fg:unconstrained-comparison}a the approximate parameters from the original demonstration of FMI \cite{hardman_time--flight_2016}, which used a condensed source and where the maximum collection efficiency was $\eta = 0.05$, and for expansion times of \SI{530}{\milli\second} and \SI{750}{\milli\second} (not shown) we find that our model agrees with the conclusions of that research, which is that FMI is the superior imaging choice when using optimized photodetectors.

Figure~\ref{fg:unconstrained-comparison}b illustrates the effect of collection efficiency $\eta$ on the choice of detection method for $N = 10^5$ atoms.  For condensed samples, we again have broad regions of $T_y$ and $\eta$ where FMI has better SNR than FI; notably, FMI is significantly better than FI as the collection efficiency decreases since FMI only needs to collect light in the same spatial mode as the incident laser.  This behaviour is also evident for laser cooled sources, where we have shown the approximate parameters from a recent study which used FMI \cite{saywell_enhancing_2023}.  The insensitivity of FMI to collection efficiency may be of particular interest for field-deployable devices where the large vacuum windows necessary for large $\eta$ add unwanted size and weight.  However, we note that large collection efficiency is not necessarily incompatible with field-deployable devices, with one recent example showcasing a collection efficiency of $\eta = 0.127$ \cite{li_continuous_2023}.  In this work, FI was used to detect $10^6$ laser-cooled atoms, which, according to our model, has significantly higher SNR than FMI for the same conditions.

Finally, Fig.~\ref{fg:unconstrained-comparison}c shows how the quality of available detectors, specifically detector NEP, affects the choice of detection method for $N = 10^5$ atoms and $\eta = 0.05$.  If the choice of detectors is limited to those with $\NEP > \SI{200}{\femto\watt/\sqrt{\hertz}}$, then FMI will provide higher detection SNR than FI.  However, if low-noise detectors are available, then laser cooled sources with relatively high collection efficiency ($\eta \sim 0.05$) will benefit from FI, while for condensed sources the optimal choice of detection method is highly dependent on the time of flight and temperature of the source.

\subsection{FMI under constraints}
\label{ssec:constrained-model}

In the previous sections, we considered the performance of FMI in the ``unconstrained'' limit where the modulation frequency is arbitrarily large and sufficient optical power is available such that the detector is shot noise limited for any first order saturation parameter $s_1$.  In realistic CAIs, there is usually a limit to the amount of optical power that can be devoted to detection, and phase/frequency modulators have limited modulation bandwidths and have damage thresholds which limit the optical power.    Under these two constraints, modulation frequency and total optical power, the optimal choice of modulation depth, and therefore first order saturation parameter, may be different due to the effect of additional modulation orders on the signal.

We consider here a generalised version of the model of FMI presented in Sec.~\ref{ssec:fmi-theory} where the amplitude of the incident electric field is time-dependent
\begin{equation}
	E(\vr,t) = \sum_{n = -\infty}^\infty E_n(\vr)e^{i n \omega_m t}
	\label{eq:phase-modulated-electric-field}
\end{equation} 
with $E_n(\vr)\propto i^nJ_n(\beta)$ for Bessel function of the first kind $J_n(x)$ and modulation depth $\beta$ according to the Jacobi-Anger expansion.  We solve Eq.~\eqref{eq:time-evolution} by expanding the Rabi frequency, coherence, and population imbalance into components oscillating at $n\omega_m$
\begin{subequations}\label{eq:temporal-expansion}
	\begin{align}
		\Omega(t) &= \sum_{n = -\infty}^\infty \Omega_n e^{in\omega_m t}\label{eq:rabi-expansion}\\
		\coh(t) &= \sum_{n = -\infty}^\infty \rho_{eg,n}(t) e^{in\omega_m t}\label{eq:coherence-expansion}\\
		w(t) &= \sum_{n = -\infty}^\infty w_n(t) e^{in\omega_m t}\label{eq:imbalance-expansion}
	\end{align}
\end{subequations}
where $\rho_{eg,n}(t)$ and $w_n(t)$ vary slowly compared to $\omega_m$.  Taking time derivatives of Eqs.~\eqref{eq:coherence-expansion} and \eqref{eq:imbalance-expansion}, substituting them into Eq.~\eqref{eq:time-evolution}, and collecting terms oscillating at the same frequency, we obtain the dynamical equations for the slowly-varying components
\begin{subequations}\label{eq:component-evolution}
	\begin{align}
		\dot{\rho}_{eg,n} &= i\left(\Delta + i\frac{\Gamma}{2} - n\omega_m\right)\rho_{eg,n}\notag\\
		&\quad - i\sum_{j = -\infty}^\infty \Omega_{n - j}w_j\label{eq:cn-evolution}\\
		\dot{w}_n &= -\Gamma\delta_{n0} - (in\omega_m + \Gamma)w_n\notag\\
		&\quad + 2i\sum_{j = -\infty}^\infty(\Omega_{n+j}\rho_{eg,j}^* - \Omega_{j-n}^*\rho_{eg,j})\label{eq:wn-evolution}
	\end{align}
\end{subequations}
where $\delta_{nm}$ is the Kronecker delta.  Note that, since population imbalances must be real quantities, $w_n = w_{-n}^*$; however, there are no such restrictions on the coherences $\rho_{eg,n}$.

Similarly to Sec.~\ref{sec:theory}, we solve Eq.~\eqref{eq:component-evolution} in the steady state by assuming that the atom-light interaction time is much longer than $\Gamma^{-1}$ so that all time derivatives are zero.  Equation~\eqref{eq:component-evolution} can then be solved for the components by converting it into a matrix equation of the form $M\mathbf{v} = \mathbf{b}$, where $\mathbf{v}$ consists of the components $w_n$, $\rho_{eg,n}$, and $\rho_{eg,n}^*$ up to a maximum order $n_{\rm max}$, and the elements of $\mathbf{b}$ are zero except for the element corresponding to $w_0$, which is $\Gamma$.  While rigourous, this method is computationally expensive, and a simpler solution suffices for determining the FMI signal and SNR in typical situations.  We note that, since $\Delta E_n \propto \rho_{eg,n}$, terms proportional to $w_{n \neq 0}$ introduce coupling between different modulation orders $E_n$ and $E_{n' \neq n}$, and these terms are, to leading order, proportional to $\Omega_n\Omega_0/(\Gamma\omega_m)$.  When both the total intensity and the modulation depth are low, so that $\Omega_n\Omega_0/(\Gamma\omega_m) \ll 1$, these nonlinear inter-modulation effects will be small, and we can neglect terms proportional to $w_{m \neq 0}$.  This allows us to solve Eq.~\eqref{eq:component-evolution} for the steady state values of $\rho_{eg,n}$ and $w_0$
\begin{subequations}\label{eq:component-steady-state}
	\begin{align}
		\rho_{eg,n} &= \frac{\Omega_n w_0}{\Delta - n\omega_m + i\frac{\Gamma}{2}}\label{eq:cn-steady-state}\\
		w_0 &= \left(1 + \sum_{j=-\infty}^\infty \frac{2|\Omega_j|^2}{(\Delta - j\omega_m)^2 + \Gamma^2/4}\right)^{-1}\label{eq:wn-steady-state}\\
		&= \left(1 + \sum_{j = -\infty}^\infty s_j\right)^{-1}
	\end{align}
\end{subequations}
where $s_j$ is the saturation parameter associated with the $j$-th sideband.  By neglecting inter-modulation effects, we have reduced the action of the additional sidebands to their off-resonant effect on the saturation of the transition.  Similarly to Eq.~\eqref{eq:electric-field-change}, we now have $\Delta E_n = i\frac{\hbar\Gamma}{4d}\OD \rho_{eg,n}$, and the intensity of the field oscillating at $n\omega_m$ is
\begin{equation}
	I_{n\omega_m}(x,y,t) = 4\epsilon_0 c\left|\sum_{j = -\infty}^\infty E_j E^*_{j - n}\right|\cos(n\omega_m t + \phi)
	\label{eq:intensity-components}
\end{equation}
where
\begin{align}
	\sum_{j = -\infty}^\infty E_j E^*_{j - n} &= \sum_{j = -\infty}^\infty\left(E_j\Delta E^*_{j - n} + E_{j - n}^*\Delta E_j\right.\notag\\
	&\quad \left. + \Delta E_j \Delta E_{j - n}^*\right).
\end{align}
The maximum signal at $n\omega_m$ is then simply the amplitude of Eq.~\eqref{eq:intensity-components}.

\begin{figure}[tb]
	\centering
	\includegraphics[width=\columnwidth]{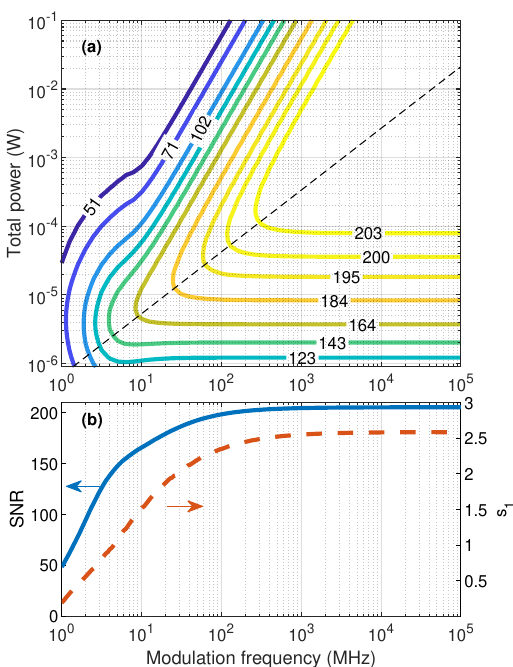}
	\caption{Optimised peak SNR for FMI under constraints.  \textbf{(a)} Contour plot (solid lines) of peak SNR for FMI with $N = 10^5$ \Rb{} atoms, $T_x = T_y = \SI{50}{\nano\kelvin}$, $t_{\rm tof} = \SI{200}{\milli\second}$, and detector noise parameters consistent with the Thorlabs DET02A detector.  \textbf{(b)} Optimal SNR as a function of modulation frequency corresponding to the black dashed line in \textbf{(a)} (blue solid, left axis) and optimal first-order saturation parameter (red dashed, right axis).}
	\label{fg:constrained-fmi-snr}
\end{figure}
Figure \ref{fg:constrained-fmi-snr}a plots contours of the peak SNR for FMI under the conditions of $N = 10^5$ \Rb{} atoms, $T_x = T_y = \SI{50}{\nano\kelvin}$, $t_{\rm tof} = \SI{200}{\milli\second}$, and detector noise parameters consistent with the Thorlabs DET02A detector.  Using the steady state solution given by Eq.~\eqref{eq:component-steady-state} up to $n_{\rm max} = 5$ and an intensity calculated using Eq.~\eqref{eq:intensity-components}, we optimise the SNR at each total optical power and modulation frequency by adjusting the modulation depth $\beta$.  We have verified that the difference between using the simplified model of Eq.~\eqref{eq:component-steady-state} and the full solution of Eq.~\eqref{eq:component-evolution} is negligible.  Figure~\ref{fg:constrained-fmi-snr}b shows the optimal SNR and corresponding first-order saturation parameter corresponding to a particular cut through the contour plot.  We see that FMI reaches its maximum SNR at relatively modest modulation frequencies of \SI{300}{\mega\hertz} and optical powers of \SI{100}{\micro\watt}, both of which are compatible with standard fibre-coupled electro-optic phase modulators in the near-infrared.  Excellent performance of $90\%$ of the maximum SNR ($184/203$) can still be achieved by reducing the modulation frequency and the optical power by an order of magnitude to \SI{30}{\mega\hertz} and \SI{10}{\micro\watt}, respectively.  The optimum modulation depth at these parameters is only $\beta = 0.35$, which puts $3\%$ of the optical power into the resonant first order sideband and less than $0.05\%$ into sidebands other than $0,\pm 1$.  In multi-level atoms, care needs to be taken to avoid placing higher-order sidebands $E_{n > 1}$ with detunings $-n\omega_m$ on-resonance with other excited states.  Small deviations in the laser polarization from perfect $\sigma^+$ will allow these sidebands to couple to the other excited states, which will lead to rapid optical pumping to the dark ground-state manifold.  Optical pumping on the timescale of the transit of the atomic cloud through the beam can be avoided with small changes of the modulation frequency so that these sidebands are detuned from the undesired transitions by several linewidths.

\subsection{Detection of squeezing}
\label{ssec:squeezing}

\begin{figure}[t]
	\centering
	\includegraphics[width=\columnwidth]{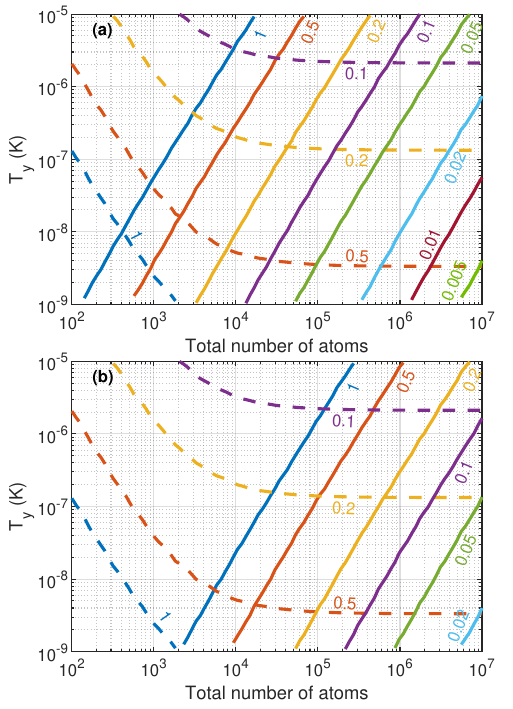}
	\caption{Contours of minimum detectable squeezing $\xi_{\rm min}$ for \Rb{} atoms using FMI (solid lines) and FI (dashed lines) with an optimised choice of photodetectors and $\eta = 0.05$ for \textbf{(a)} $t_{\rm tof} = \SI{50}{\milli\second}$ and \textbf{(b)} $t_{\rm tof} = \SI{200}{\milli\second}$.}
	\label{fg:squeezing}
\end{figure}

Next-generation CAIs aim to generate entangled atomic sources that exhibit squeezing of the phase noise below the atom number shot noise limit \cite{szigeti_high-precision_2020,corgier_delta-kick_2021,kritsotakis_spin_2021,szigeti_improving_2021}, which will enable higher-precision measurements.  Through either a quantum non-demolition measurement \cite{kritsotakis_spin_2021} or atomic interactions \cite{szigeti_high-precision_2020}, an entangled atomic source can be prepared where the fluctuations in the population difference between two states, $N_1 - N_2$, is below the atom number shot noise limit.  Specifically, the variance $\mathrm{Var}(N_1 - N_2) \approx \xi^2 C^2N$ for Wineland spin-squeezing parameter $\xi$ \cite{wineland_squeezed_1994} and with interferometer contrast $C$ and total atom number $N = N_1 + N_2$.  The variance in the measured phase $\phi$ at mid-fringe ($N_1 \approx N_2 \approx N/2$) is then
\begin{equation}
	\textrm{Var}(\phi) = \frac{1}{N}\left[\xi^2 + \frac{N/2}{C^2\SNR^2_{N/2,\rm det}}\right],
	\label{eq:phase-variance}
\end{equation}
where $\SNR_{N/2,\rm det}$ is the SNR associated with measuring $N/2$ atoms using a given detection method.  In the absence of additional noise sources, a squeezed source will have a lower uncertainty in phase, yielding a more sensitive atom interferometer.  In the presence of detection noise, the minimum squeezing that can be detected is $\xi_{\rm min} = C^{-1}\SNR_{N/2,\rm det}^{-1}\sqrt{N/2}$, implying that we need $\SNR_{N/2,\rm det} > C^{-1}\sqrt{N/2}$ in order to have a phase sensitivity higher than that allowed by the atom number shot noise limit.  In the following we will assume $C = 1$.

Unsurprisingly, FMI and FI behave differently regarding their ability to detect atoms below the standard quantum limit.  In the limit where the detectors are photon shot noise limited, FMI has an SNR of $\SNR_{N,\rm FMI} = \OD\sqrt{R_0P_{\rm sat}\tau_s/(2e)}$ where the averaging time $\tau_s$ in this instance is the effective temporal width of the sample as it passes through the detection beam.  For detection beams with $w_y \ll a_y$ and for expansion times long enough that the initial size of the sample can be neglected, the effective temporal width is $\tau_s = \sqrt{4\pi k_B T_y/(mg^2)}$ for gravitational acceleration $g$, and therefore
\begin{equation}
	\xi_{\textrm{min, FMI}} = \sqrt{\frac{2e}{\sigma I_{\rm sat} R_0 \tau_s f_x f_y \OD}}
	\label{eq:min-squeezing-fmi}
\end{equation}
where $\sigma$ is the on-resonance absorption cross-section.  In contrast, since $\SNR_{\rm FI} \propto \sqrt{\OD} \propto \sqrt{N}$ the minimum detectable squeezing for FI is
\begin{equation}
	\xi_{\textrm{min, FI}} = \sqrt{\frac{4eF}{\eta \sigma I_{\rm sat} R_0 f_x f_y \tau_s}}
	\label{eq:min-squeezing-fi}
\end{equation}
which is independent of the $\OD$ and decreases with increasing temperature; however, a minimum $\OD$ must be reached in order for FI to be photon shot noise-limited.

Figure \ref{fg:squeezing} shows contours of minimum detectable squeezing $\xi_{\rm min} = \sqrt{N/2}/\SNR_{N/2,\rm det}$ for both FMI and FI at two expansion times of $t_{\rm tof} = \SI{50}{\milli\second}$ and $t_{\rm tof} = \SI{200}{\milli\second}$.  As expected, for FMI the minimum detectable squeezing worsens for lower atom number and higher temperature, whereas for FI $\xi_{\rm min}$ is independent of $N$ for large numbers.  Furthermore, since the power scattered in FI is proportional to $\OD P_{\rm sat}$, it is independent of $t_{\rm tof}$, and thus so is $\SNR_{N/2,\rm FI}$.  As seen in both Eq.~\eqref{eq:min-squeezing-fi} and Fig.~\ref{fg:squeezing}, the minimum measurable squeezing for FI is therefore independent of $t_{\rm tof}$.  In contrast, $\SNR_{N,\rm FMI}$ depends strongly on $t_{\rm tof}$ since it is proportional to $\sqrt{\OD}$.  When $t_{\rm tof}$ is small, such as in Fig.~\ref{fg:squeezing}a, FMI achieves a smaller $\xi_{\rm min}$ than FI for $N > 10^5$, whereas for longer $t_{\rm tof}$ such as in Fig.~\ref{fg:squeezing}b FI is the better option for $N \lessapprox 10^5$ and is the only viable option for $N < 10^4$.
\section{Conclusion}
\label{sec:conclusion}

In this article, we have presented a thorough analysis of time-of-flight measurements using frequency modulation imaging or fluorescence imaging for measuring atom density and number in cold atom interferometers.  We have presented strategies for optimising the SNR for FMI, showing a clear optimum choice for horizontal beam waist and saturation parameter, as well as demonstrating that only modest modulation frequencies and total beam powers are needed to reach ideal SNRs.  We have investigated under what parameter regimes FMI is the superior detection strategy compared to FI and have found that FMI is the preferred method for condensed atomic sources in compact interferometers or when design considerations significantly restrict the collection efficiency.  Finally, we have shown that next-generation CAIs using squeezed atomic sources may benefit from using FMI for compact interferometers or for large atom number.

\begin{acknowledgements}

The authors were supported through Australian Research Council grants FT210100809 (S.A.H.), DP190101709 (R.J.T.), and LP190100621 (S.L. and R.J.T.).

\end{acknowledgements}

\bibliographystyle{apsrev4-2}

\begin{thebibliography}{35}%
\makeatletter
\providecommand \@ifxundefined [1]{%
 \@ifx{#1\undefined}
}%
\providecommand \@ifnum [1]{%
 \ifnum #1\expandafter \@firstoftwo
 \else \expandafter \@secondoftwo
 \fi
}%
\providecommand \@ifx [1]{%
 \ifx #1\expandafter \@firstoftwo
 \else \expandafter \@secondoftwo
 \fi
}%
\providecommand \natexlab [1]{#1}%
\providecommand \enquote  [1]{``#1''}%
\providecommand \bibnamefont  [1]{#1}%
\providecommand \bibfnamefont [1]{#1}%
\providecommand \citenamefont [1]{#1}%
\providecommand \href@noop [0]{\@secondoftwo}%
\providecommand \href [0]{\begingroup \@sanitize@url \@href}%
\providecommand \@href[1]{\@@startlink{#1}\@@href}%
\providecommand \@@href[1]{\endgroup#1\@@endlink}%
\providecommand \@sanitize@url [0]{\catcode `\\12\catcode `\$12\catcode
  `\&12\catcode `\#12\catcode `\^12\catcode `\_12\catcode `\%12\relax}%
\providecommand \@@startlink[1]{}%
\providecommand \@@endlink[0]{}%
\providecommand \url  [0]{\begingroup\@sanitize@url \@url }%
\providecommand \@url [1]{\endgroup\@href {#1}{\urlprefix }}%
\providecommand \urlprefix  [0]{URL }%
\providecommand \Eprint [0]{\href }%
\providecommand \doibase [0]{https://doi.org/}%
\providecommand \selectlanguage [0]{\@gobble}%
\providecommand \bibinfo  [0]{\@secondoftwo}%
\providecommand \bibfield  [0]{\@secondoftwo}%
\providecommand \translation [1]{[#1]}%
\providecommand \BibitemOpen [0]{}%
\providecommand \bibitemStop [0]{}%
\providecommand \bibitemNoStop [0]{.\EOS\space}%
\providecommand \EOS [0]{\spacefactor3000\relax}%
\providecommand \BibitemShut  [1]{\csname bibitem#1\endcsname}%
\let\auto@bib@innerbib\@empty
\bibitem [{\citenamefont {Asenbaum}\ \emph {et~al.}(2020)\citenamefont
  {Asenbaum}, \citenamefont {Overstreet}, \citenamefont {Kim}, \citenamefont
  {Curti},\ and\ \citenamefont
  {Kasevich}}]{asenbaum_atom-interferometric_2020}%
  \BibitemOpen
  \bibfield  {author} {\bibinfo {author} {\bibfnamefont {P.}~\bibnamefont
  {Asenbaum}}, \bibinfo {author} {\bibfnamefont {C.}~\bibnamefont
  {Overstreet}}, \bibinfo {author} {\bibfnamefont {M.}~\bibnamefont {Kim}},
  \bibinfo {author} {\bibfnamefont {J.}~\bibnamefont {Curti}},\ and\ \bibinfo
  {author} {\bibfnamefont {M.~A.}\ \bibnamefont {Kasevich}},\ }\href
  {https://doi.org/10.1103/PhysRevLett.125.191101} {\bibfield  {journal}
  {\bibinfo  {journal} {Physical Review Letters}\ }\textbf {\bibinfo {volume}
  {125}},\ \bibinfo {pages} {191101} (\bibinfo {year} {2020})}\BibitemShut
  {NoStop}%
\bibitem [{\citenamefont {Fixler}\ \emph {et~al.}(2007)\citenamefont {Fixler},
  \citenamefont {Foster}, \citenamefont {McGuirk},\ and\ \citenamefont
  {Kasevich}}]{fixler_atom_2007}%
  \BibitemOpen
  \bibfield  {author} {\bibinfo {author} {\bibfnamefont {J.~B.}\ \bibnamefont
  {Fixler}}, \bibinfo {author} {\bibfnamefont {G.~T.}\ \bibnamefont {Foster}},
  \bibinfo {author} {\bibfnamefont {J.~M.}\ \bibnamefont {McGuirk}},\ and\
  \bibinfo {author} {\bibfnamefont {M.~A.}\ \bibnamefont {Kasevich}},\ }\href
  {https://doi.org/10.1126/science.1135459} {\bibfield  {journal} {\bibinfo
  {journal} {Science}\ }\textbf {\bibinfo {volume} {315}},\ \bibinfo {pages}
  {74} (\bibinfo {year} {2007})}\BibitemShut {NoStop}%
\bibitem [{\citenamefont {Rosi}\ \emph {et~al.}(2014)\citenamefont {Rosi},
  \citenamefont {Sorrentino}, \citenamefont {Cacciapuoti}, \citenamefont
  {Prevedelli},\ and\ \citenamefont {Tino}}]{rosi_precision_2014}%
  \BibitemOpen
  \bibfield  {author} {\bibinfo {author} {\bibfnamefont {G.}~\bibnamefont
  {Rosi}}, \bibinfo {author} {\bibfnamefont {F.}~\bibnamefont {Sorrentino}},
  \bibinfo {author} {\bibfnamefont {L.}~\bibnamefont {Cacciapuoti}}, \bibinfo
  {author} {\bibfnamefont {M.}~\bibnamefont {Prevedelli}},\ and\ \bibinfo
  {author} {\bibfnamefont {G.~M.}\ \bibnamefont {Tino}},\ }\href
  {https://doi.org/10.1038/nature13433} {\bibfield  {journal} {\bibinfo
  {journal} {Nature}\ }\textbf {\bibinfo {volume} {510}},\ \bibinfo {pages}
  {518} (\bibinfo {year} {2014})},\ \bibinfo {note} {number: 7506 Publisher:
  Nature Publishing Group}\BibitemShut {NoStop}%
\bibitem [{\citenamefont {Weiss}\ \emph {et~al.}(1994)\citenamefont {Weiss},
  \citenamefont {Young},\ and\ \citenamefont {Chu}}]{weiss_precision_1994}%
  \BibitemOpen
  \bibfield  {author} {\bibinfo {author} {\bibfnamefont {D.~S.}\ \bibnamefont
  {Weiss}}, \bibinfo {author} {\bibfnamefont {B.~C.}\ \bibnamefont {Young}},\
  and\ \bibinfo {author} {\bibfnamefont {S.}~\bibnamefont {Chu}},\ }\href
  {https://doi.org/10.1007/BF01081393} {\bibfield  {journal} {\bibinfo
  {journal} {Applied Physics B}\ }\textbf {\bibinfo {volume} {59}},\ \bibinfo
  {pages} {217} (\bibinfo {year} {1994})}\BibitemShut {NoStop}%
\bibitem [{\citenamefont {Lan}\ \emph {et~al.}(2013)\citenamefont {Lan},
  \citenamefont {Kuan}, \citenamefont {Estey}, \citenamefont {English},
  \citenamefont {Brown}, \citenamefont {Hohensee},\ and\ \citenamefont
  {Müller}}]{lan_clock_2013}%
  \BibitemOpen
  \bibfield  {author} {\bibinfo {author} {\bibfnamefont {S.-Y.}\ \bibnamefont
  {Lan}}, \bibinfo {author} {\bibfnamefont {P.-C.}\ \bibnamefont {Kuan}},
  \bibinfo {author} {\bibfnamefont {B.}~\bibnamefont {Estey}}, \bibinfo
  {author} {\bibfnamefont {D.}~\bibnamefont {English}}, \bibinfo {author}
  {\bibfnamefont {J.~M.}\ \bibnamefont {Brown}}, \bibinfo {author}
  {\bibfnamefont {M.~A.}\ \bibnamefont {Hohensee}},\ and\ \bibinfo {author}
  {\bibfnamefont {H.}~\bibnamefont {Müller}},\ }\href
  {https://doi.org/10.1126/science.1230767} {\bibfield  {journal} {\bibinfo
  {journal} {Science}\ }\textbf {\bibinfo {volume} {339}},\ \bibinfo {pages}
  {554} (\bibinfo {year} {2013})}\BibitemShut {NoStop}%
\bibitem [{\citenamefont {Yu}\ \emph {et~al.}(2019)\citenamefont {Yu},
  \citenamefont {Zhong}, \citenamefont {Estey}, \citenamefont {Kwan},
  \citenamefont {Parker},\ and\ \citenamefont
  {Müller}}]{yu_atom-interferometry_2019}%
  \BibitemOpen
  \bibfield  {author} {\bibinfo {author} {\bibfnamefont {C.}~\bibnamefont
  {Yu}}, \bibinfo {author} {\bibfnamefont {W.}~\bibnamefont {Zhong}}, \bibinfo
  {author} {\bibfnamefont {B.}~\bibnamefont {Estey}}, \bibinfo {author}
  {\bibfnamefont {J.}~\bibnamefont {Kwan}}, \bibinfo {author} {\bibfnamefont
  {R.~H.}\ \bibnamefont {Parker}},\ and\ \bibinfo {author} {\bibfnamefont
  {H.}~\bibnamefont {Müller}},\ }\href
  {https://doi.org/10.1002/andp.201800346} {\bibfield  {journal} {\bibinfo
  {journal} {Annalen der Physik}\ }\textbf {\bibinfo {volume} {531}},\ \bibinfo
  {pages} {1800346} (\bibinfo {year} {2019})}\BibitemShut {NoStop}%
\bibitem [{\citenamefont {Fallon}\ \emph {et~al.}(2022)\citenamefont {Fallon},
  \citenamefont {Moan}, \citenamefont {Larson},\ and\ \citenamefont
  {Sackett}}]{fallon_measurement_2022}%
  \BibitemOpen
  \bibfield  {author} {\bibinfo {author} {\bibfnamefont {A.~J.}\ \bibnamefont
  {Fallon}}, \bibinfo {author} {\bibfnamefont {E.~R.}\ \bibnamefont {Moan}},
  \bibinfo {author} {\bibfnamefont {E.~A.}\ \bibnamefont {Larson}},\ and\
  \bibinfo {author} {\bibfnamefont {C.~A.}\ \bibnamefont {Sackett}},\ }\href
  {https://doi.org/10.1103/PhysRevA.105.L030802} {\bibfield  {journal}
  {\bibinfo  {journal} {Physical Review A}\ }\textbf {\bibinfo {volume}
  {105}},\ \bibinfo {pages} {L030802} (\bibinfo {year} {2022})}\BibitemShut
  {NoStop}%
\bibitem [{\citenamefont {Leonard}\ \emph {et~al.}(2015)\citenamefont
  {Leonard}, \citenamefont {Fallon}, \citenamefont {Sackett},\ and\
  \citenamefont {Safronova}}]{leonard_high-precision_2015}%
  \BibitemOpen
  \bibfield  {author} {\bibinfo {author} {\bibfnamefont {R.~H.}\ \bibnamefont
  {Leonard}}, \bibinfo {author} {\bibfnamefont {A.~J.}\ \bibnamefont {Fallon}},
  \bibinfo {author} {\bibfnamefont {C.~A.}\ \bibnamefont {Sackett}},\ and\
  \bibinfo {author} {\bibfnamefont {M.~S.}\ \bibnamefont {Safronova}},\ }\href
  {https://doi.org/10.1103/PhysRevA.92.052501} {\bibfield  {journal} {\bibinfo
  {journal} {Physical Review A}\ }\textbf {\bibinfo {volume} {92}},\ \bibinfo
  {pages} {052501} (\bibinfo {year} {2015})}\BibitemShut {NoStop}%
\bibitem [{\citenamefont {Freier}\ \emph {et~al.}(2016)\citenamefont {Freier},
  \citenamefont {Hauth}, \citenamefont {Schkolnik}, \citenamefont {Leykauf},
  \citenamefont {Schilling}, \citenamefont {Wziontek}, \citenamefont
  {Scherneck}, \citenamefont {Müller},\ and\ \citenamefont
  {Peters}}]{freier_mobile_2016}%
  \BibitemOpen
  \bibfield  {author} {\bibinfo {author} {\bibfnamefont {C.}~\bibnamefont
  {Freier}}, \bibinfo {author} {\bibfnamefont {M.}~\bibnamefont {Hauth}},
  \bibinfo {author} {\bibfnamefont {V.}~\bibnamefont {Schkolnik}}, \bibinfo
  {author} {\bibfnamefont {B.}~\bibnamefont {Leykauf}}, \bibinfo {author}
  {\bibfnamefont {M.}~\bibnamefont {Schilling}}, \bibinfo {author}
  {\bibfnamefont {H.}~\bibnamefont {Wziontek}}, \bibinfo {author}
  {\bibfnamefont {H.-G.}\ \bibnamefont {Scherneck}}, \bibinfo {author}
  {\bibfnamefont {J.}~\bibnamefont {Müller}},\ and\ \bibinfo {author}
  {\bibfnamefont {A.}~\bibnamefont {Peters}},\ }\href
  {https://doi.org/10.1088/1742-6596/723/1/012050} {\bibfield  {journal}
  {\bibinfo  {journal} {Journal of Physics: Conference Series}\ }\textbf
  {\bibinfo {volume} {723}},\ \bibinfo {pages} {012050} (\bibinfo {year}
  {2016})}\BibitemShut {NoStop}%
\bibitem [{\citenamefont {McGuirk}\ \emph {et~al.}(2002)\citenamefont
  {McGuirk}, \citenamefont {Foster}, \citenamefont {Fixler}, \citenamefont
  {Snadden},\ and\ \citenamefont {Kasevich}}]{mcguirk_sensitive_2002}%
  \BibitemOpen
  \bibfield  {author} {\bibinfo {author} {\bibfnamefont {J.~M.}\ \bibnamefont
  {McGuirk}}, \bibinfo {author} {\bibfnamefont {G.~T.}\ \bibnamefont {Foster}},
  \bibinfo {author} {\bibfnamefont {J.~B.}\ \bibnamefont {Fixler}}, \bibinfo
  {author} {\bibfnamefont {M.~J.}\ \bibnamefont {Snadden}},\ and\ \bibinfo
  {author} {\bibfnamefont {M.~A.}\ \bibnamefont {Kasevich}},\ }\href
  {https://doi.org/10.1103/PhysRevA.65.033608} {\bibfield  {journal} {\bibinfo
  {journal} {Physical Review A}\ }\textbf {\bibinfo {volume} {65}},\ \bibinfo
  {pages} {033608} (\bibinfo {year} {2002})}\BibitemShut {NoStop}%
\bibitem [{\citenamefont {Stray}\ \emph {et~al.}(2022)\citenamefont {Stray},
  \citenamefont {Lamb}, \citenamefont {Kaushik}, \citenamefont {Vovrosh},
  \citenamefont {Rodgers}, \citenamefont {Winch}, \citenamefont {Hayati},
  \citenamefont {Boddice}, \citenamefont {Stabrawa}, \citenamefont {Niggebaum},
  \citenamefont {Langlois}, \citenamefont {Lien}, \citenamefont {Lellouch},
  \citenamefont {Roshanmanesh}, \citenamefont {Ridley}, \citenamefont
  {de~Villiers}, \citenamefont {Brown}, \citenamefont {Cross}, \citenamefont
  {Tuckwell}, \citenamefont {Faramarzi}, \citenamefont {Metje}, \citenamefont
  {Bongs},\ and\ \citenamefont {Holynski}}]{stray_quantum_2022}%
  \BibitemOpen
  \bibfield  {author} {\bibinfo {author} {\bibfnamefont {B.}~\bibnamefont
  {Stray}}, \bibinfo {author} {\bibfnamefont {A.}~\bibnamefont {Lamb}},
  \bibinfo {author} {\bibfnamefont {A.}~\bibnamefont {Kaushik}}, \bibinfo
  {author} {\bibfnamefont {J.}~\bibnamefont {Vovrosh}}, \bibinfo {author}
  {\bibfnamefont {A.}~\bibnamefont {Rodgers}}, \bibinfo {author} {\bibfnamefont
  {J.}~\bibnamefont {Winch}}, \bibinfo {author} {\bibfnamefont
  {F.}~\bibnamefont {Hayati}}, \bibinfo {author} {\bibfnamefont
  {D.}~\bibnamefont {Boddice}}, \bibinfo {author} {\bibfnamefont
  {A.}~\bibnamefont {Stabrawa}}, \bibinfo {author} {\bibfnamefont
  {A.}~\bibnamefont {Niggebaum}}, \bibinfo {author} {\bibfnamefont
  {M.}~\bibnamefont {Langlois}}, \bibinfo {author} {\bibfnamefont {Y.-H.}\
  \bibnamefont {Lien}}, \bibinfo {author} {\bibfnamefont {S.}~\bibnamefont
  {Lellouch}}, \bibinfo {author} {\bibfnamefont {S.}~\bibnamefont
  {Roshanmanesh}}, \bibinfo {author} {\bibfnamefont {K.}~\bibnamefont
  {Ridley}}, \bibinfo {author} {\bibfnamefont {G.}~\bibnamefont {de~Villiers}},
  \bibinfo {author} {\bibfnamefont {G.}~\bibnamefont {Brown}}, \bibinfo
  {author} {\bibfnamefont {T.}~\bibnamefont {Cross}}, \bibinfo {author}
  {\bibfnamefont {G.}~\bibnamefont {Tuckwell}}, \bibinfo {author}
  {\bibfnamefont {A.}~\bibnamefont {Faramarzi}}, \bibinfo {author}
  {\bibfnamefont {N.}~\bibnamefont {Metje}}, \bibinfo {author} {\bibfnamefont
  {K.}~\bibnamefont {Bongs}},\ and\ \bibinfo {author} {\bibfnamefont
  {M.}~\bibnamefont {Holynski}},\ }\href
  {https://doi.org/10.1038/s41586-021-04315-3} {\bibfield  {journal} {\bibinfo
  {journal} {Nature}\ }\textbf {\bibinfo {volume} {602}},\ \bibinfo {pages}
  {590} (\bibinfo {year} {2022})},\ \bibinfo {note} {number: 7898 Publisher:
  Nature Publishing Group}\BibitemShut {NoStop}%
\bibitem [{\citenamefont {Canuel}\ \emph {et~al.}(2006)\citenamefont {Canuel},
  \citenamefont {Leduc}, \citenamefont {Holleville}, \citenamefont {Gauguet},
  \citenamefont {Fils}, \citenamefont {Virdis}, \citenamefont {Clairon},
  \citenamefont {Dimarcq}, \citenamefont {Bord{\'e}}, \citenamefont
  {Landragin},\ and\ \citenamefont {Bouyer}}]{canuel_six-axis_2006}%
  \BibitemOpen
  \bibfield  {author} {\bibinfo {author} {\bibfnamefont {B.}~\bibnamefont
  {Canuel}}, \bibinfo {author} {\bibfnamefont {F.}~\bibnamefont {Leduc}},
  \bibinfo {author} {\bibfnamefont {D.}~\bibnamefont {Holleville}}, \bibinfo
  {author} {\bibfnamefont {A.}~\bibnamefont {Gauguet}}, \bibinfo {author}
  {\bibfnamefont {J.}~\bibnamefont {Fils}}, \bibinfo {author} {\bibfnamefont
  {A.}~\bibnamefont {Virdis}}, \bibinfo {author} {\bibfnamefont
  {A.}~\bibnamefont {Clairon}}, \bibinfo {author} {\bibfnamefont
  {N.}~\bibnamefont {Dimarcq}}, \bibinfo {author} {\bibfnamefont {C.~J.}\
  \bibnamefont {Bord{\'e}}}, \bibinfo {author} {\bibfnamefont {A.}~\bibnamefont
  {Landragin}},\ and\ \bibinfo {author} {\bibfnamefont {P.}~\bibnamefont
  {Bouyer}},\ }\href {https://doi.org/10.1103/PhysRevLett.97.010402} {\bibfield
   {journal} {\bibinfo  {journal} {Physical Review Letters}\ }\textbf {\bibinfo
  {volume} {97}},\ \bibinfo {pages} {010402} (\bibinfo {year}
  {2006})}\BibitemShut {NoStop}%
\bibitem [{\citenamefont {Berg}\ \emph {et~al.}(2015)\citenamefont {Berg},
  \citenamefont {Abend}, \citenamefont {Tackmann}, \citenamefont {Schubert},
  \citenamefont {Giese}, \citenamefont {Schleich}, \citenamefont {Narducci},
  \citenamefont {Ertmer},\ and\ \citenamefont
  {Rasel}}]{berg_composite-light-pulse_2015}%
  \BibitemOpen
  \bibfield  {author} {\bibinfo {author} {\bibfnamefont {P.}~\bibnamefont
  {Berg}}, \bibinfo {author} {\bibfnamefont {S.}~\bibnamefont {Abend}},
  \bibinfo {author} {\bibfnamefont {G.}~\bibnamefont {Tackmann}}, \bibinfo
  {author} {\bibfnamefont {C.}~\bibnamefont {Schubert}}, \bibinfo {author}
  {\bibfnamefont {E.}~\bibnamefont {Giese}}, \bibinfo {author} {\bibfnamefont
  {W.~P.}\ \bibnamefont {Schleich}}, \bibinfo {author} {\bibfnamefont {F.~A.}\
  \bibnamefont {Narducci}}, \bibinfo {author} {\bibfnamefont {W.}~\bibnamefont
  {Ertmer}},\ and\ \bibinfo {author} {\bibfnamefont {E.~M.}\ \bibnamefont
  {Rasel}},\ }\href {https://doi.org/10.1103/PhysRevLett.114.063002} {\bibfield
   {journal} {\bibinfo  {journal} {Physical Review Letters}\ }\textbf {\bibinfo
  {volume} {114}},\ \bibinfo {pages} {063002} (\bibinfo {year}
  {2015})}\BibitemShut {NoStop}%
\bibitem [{\citenamefont {Szigeti}\ \emph {et~al.}(2020)\citenamefont
  {Szigeti}, \citenamefont {Nolan}, \citenamefont {Close},\ and\ \citenamefont
  {Haine}}]{szigeti_high-precision_2020}%
  \BibitemOpen
  \bibfield  {author} {\bibinfo {author} {\bibfnamefont {S.~S.}\ \bibnamefont
  {Szigeti}}, \bibinfo {author} {\bibfnamefont {S.~P.}\ \bibnamefont {Nolan}},
  \bibinfo {author} {\bibfnamefont {J.~D.}\ \bibnamefont {Close}},\ and\
  \bibinfo {author} {\bibfnamefont {S.~A.}\ \bibnamefont {Haine}},\ }\href
  {https://doi.org/10.1103/PhysRevLett.125.100402} {\bibfield  {journal}
  {\bibinfo  {journal} {Physical Review Letters}\ }\textbf {\bibinfo {volume}
  {125}},\ \bibinfo {pages} {100402} (\bibinfo {year} {2020})}\BibitemShut
  {NoStop}%
\bibitem [{\citenamefont {Corgier}\ \emph {et~al.}(2021)\citenamefont
  {Corgier}, \citenamefont {Gaaloul}, \citenamefont {Smerzi},\ and\
  \citenamefont {Pezz{\`e}}}]{corgier_delta-kick_2021}%
  \BibitemOpen
  \bibfield  {author} {\bibinfo {author} {\bibfnamefont {R.}~\bibnamefont
  {Corgier}}, \bibinfo {author} {\bibfnamefont {N.}~\bibnamefont {Gaaloul}},
  \bibinfo {author} {\bibfnamefont {A.}~\bibnamefont {Smerzi}},\ and\ \bibinfo
  {author} {\bibfnamefont {L.}~\bibnamefont {Pezz{\`e}}},\ }\href
  {https://doi.org/10.1103/PhysRevLett.127.183401} {\bibfield  {journal}
  {\bibinfo  {journal} {Physical Review Letters}\ }\textbf {\bibinfo {volume}
  {127}},\ \bibinfo {pages} {183401} (\bibinfo {year} {2021})}\BibitemShut
  {NoStop}%
\bibitem [{\citenamefont {Kritsotakis}\ \emph {et~al.}(2021)\citenamefont
  {Kritsotakis}, \citenamefont {Dunningham},\ and\ \citenamefont
  {Haine}}]{kritsotakis_spin_2021}%
  \BibitemOpen
  \bibfield  {author} {\bibinfo {author} {\bibfnamefont {M.}~\bibnamefont
  {Kritsotakis}}, \bibinfo {author} {\bibfnamefont {J.~A.}\ \bibnamefont
  {Dunningham}},\ and\ \bibinfo {author} {\bibfnamefont {S.~A.}\ \bibnamefont
  {Haine}},\ }\href {https://doi.org/10.1103/PhysRevA.103.023318} {\bibfield
  {journal} {\bibinfo  {journal} {Physical Review A}\ }\textbf {\bibinfo
  {volume} {103}},\ \bibinfo {pages} {023318} (\bibinfo {year}
  {2021})}\BibitemShut {NoStop}%
\bibitem [{\citenamefont {Hardman}\ \emph
  {et~al.}(2016{\natexlab{a}})\citenamefont {Hardman}, \citenamefont {Wigley},
  \citenamefont {Everitt}, \citenamefont {Manju}, \citenamefont {Kuhn},\ and\
  \citenamefont {Robins}}]{hardman_time--flight_2016}%
  \BibitemOpen
  \bibfield  {author} {\bibinfo {author} {\bibfnamefont {K.~S.}\ \bibnamefont
  {Hardman}}, \bibinfo {author} {\bibfnamefont {P.~B.}\ \bibnamefont {Wigley}},
  \bibinfo {author} {\bibfnamefont {P.~J.}\ \bibnamefont {Everitt}}, \bibinfo
  {author} {\bibfnamefont {P.}~\bibnamefont {Manju}}, \bibinfo {author}
  {\bibfnamefont {C.~C.~N.}\ \bibnamefont {Kuhn}},\ and\ \bibinfo {author}
  {\bibfnamefont {N.~P.}\ \bibnamefont {Robins}},\ }\href
  {https://doi.org/10.1364/OL.41.002505} {\bibfield  {journal} {\bibinfo
  {journal} {Optics Letters}\ }\textbf {\bibinfo {volume} {41}},\ \bibinfo
  {pages} {2505} (\bibinfo {year} {2016}{\natexlab{a}})}\BibitemShut {NoStop}%
\bibitem [{\citenamefont {Cheng}\ \emph {et~al.}(2018)\citenamefont {Cheng},
  \citenamefont {Zhang}, \citenamefont {Chen}, \citenamefont {Zhang},
  \citenamefont {Xu}, \citenamefont {Duan}, \citenamefont {Zhou},\ and\
  \citenamefont {Hu}}]{cheng_momentum-resolved_2018}%
  \BibitemOpen
  \bibfield  {author} {\bibinfo {author} {\bibfnamefont {Y.}~\bibnamefont
  {Cheng}}, \bibinfo {author} {\bibfnamefont {K.}~\bibnamefont {Zhang}},
  \bibinfo {author} {\bibfnamefont {L.-L.}\ \bibnamefont {Chen}}, \bibinfo
  {author} {\bibfnamefont {T.}~\bibnamefont {Zhang}}, \bibinfo {author}
  {\bibfnamefont {W.-J.}\ \bibnamefont {Xu}}, \bibinfo {author} {\bibfnamefont
  {X.-C.}\ \bibnamefont {Duan}}, \bibinfo {author} {\bibfnamefont {M.-K.}\
  \bibnamefont {Zhou}},\ and\ \bibinfo {author} {\bibfnamefont {Z.-K.}\
  \bibnamefont {Hu}},\ }\href {https://doi.org/10.1103/PhysRevA.98.043611}
  {\bibfield  {journal} {\bibinfo  {journal} {Physical Review A}\ }\textbf
  {\bibinfo {volume} {98}},\ \bibinfo {pages} {043611} (\bibinfo {year}
  {2018})}\BibitemShut {NoStop}%
\bibitem [{\citenamefont {Gauguet}\ \emph {et~al.}(2009)\citenamefont
  {Gauguet}, \citenamefont {Canuel}, \citenamefont {L{\'e}v{\`e}que},
  \citenamefont {Chaibi},\ and\ \citenamefont
  {Landragin}}]{gauguet_characterization_2009}%
  \BibitemOpen
  \bibfield  {author} {\bibinfo {author} {\bibfnamefont {A.}~\bibnamefont
  {Gauguet}}, \bibinfo {author} {\bibfnamefont {B.}~\bibnamefont {Canuel}},
  \bibinfo {author} {\bibfnamefont {T.}~\bibnamefont {L{\'e}v{\`e}que}},
  \bibinfo {author} {\bibfnamefont {W.}~\bibnamefont {Chaibi}},\ and\ \bibinfo
  {author} {\bibfnamefont {A.}~\bibnamefont {Landragin}},\ }\href
  {https://doi.org/10.1103/PhysRevA.80.063604} {\bibfield  {journal} {\bibinfo
  {journal} {Physical Review A}\ }\textbf {\bibinfo {volume} {80}},\ \bibinfo
  {pages} {063604} (\bibinfo {year} {2009})}\BibitemShut {NoStop}%
\bibitem [{\citenamefont {Wu}\ \emph {et~al.}(2019)\citenamefont {Wu},
  \citenamefont {Pagel}, \citenamefont {Malek}, \citenamefont {Nguyen},
  \citenamefont {Zi}, \citenamefont {Scheirer},\ and\ \citenamefont
  {Müller}}]{wu_gravity_2019}%
  \BibitemOpen
  \bibfield  {author} {\bibinfo {author} {\bibfnamefont {X.}~\bibnamefont
  {Wu}}, \bibinfo {author} {\bibfnamefont {Z.}~\bibnamefont {Pagel}}, \bibinfo
  {author} {\bibfnamefont {B.~S.}\ \bibnamefont {Malek}}, \bibinfo {author}
  {\bibfnamefont {T.~H.}\ \bibnamefont {Nguyen}}, \bibinfo {author}
  {\bibfnamefont {F.}~\bibnamefont {Zi}}, \bibinfo {author} {\bibfnamefont
  {D.~S.}\ \bibnamefont {Scheirer}},\ and\ \bibinfo {author} {\bibfnamefont
  {H.}~\bibnamefont {Müller}},\ }\href
  {https://doi.org/10.1126/sciadv.aax0800} {\bibfield  {journal} {\bibinfo
  {journal} {Science Advances}\ }\textbf {\bibinfo {volume} {5}},\ \bibinfo
  {pages} {eaax0800} (\bibinfo {year} {2019})},\ \bibinfo {note} {publisher:
  American Association for the Advancement of Science Section: Research
  Article}\BibitemShut {NoStop}%
\bibitem [{\citenamefont {Rudolph}\ \emph {et~al.}(2015)\citenamefont
  {Rudolph}, \citenamefont {Herr}, \citenamefont {Grzeschik}, \citenamefont
  {Sternke}, \citenamefont {Grote}, \citenamefont {Popp}, \citenamefont
  {Becker}, \citenamefont {Müntinga}, \citenamefont {Ahlers}, \citenamefont
  {Peters}, \citenamefont {Lämmerzahl}, \citenamefont {Sengstock},
  \citenamefont {Gaaloul}, \citenamefont {Ertmer},\ and\ \citenamefont
  {Rasel}}]{rudolph_high-flux_2015}%
  \BibitemOpen
  \bibfield  {author} {\bibinfo {author} {\bibfnamefont {J.}~\bibnamefont
  {Rudolph}}, \bibinfo {author} {\bibfnamefont {W.}~\bibnamefont {Herr}},
  \bibinfo {author} {\bibfnamefont {C.}~\bibnamefont {Grzeschik}}, \bibinfo
  {author} {\bibfnamefont {T.}~\bibnamefont {Sternke}}, \bibinfo {author}
  {\bibfnamefont {A.}~\bibnamefont {Grote}}, \bibinfo {author} {\bibfnamefont
  {M.}~\bibnamefont {Popp}}, \bibinfo {author} {\bibfnamefont {D.}~\bibnamefont
  {Becker}}, \bibinfo {author} {\bibfnamefont {H.}~\bibnamefont {Müntinga}},
  \bibinfo {author} {\bibfnamefont {H.}~\bibnamefont {Ahlers}}, \bibinfo
  {author} {\bibfnamefont {A.}~\bibnamefont {Peters}}, \bibinfo {author}
  {\bibfnamefont {C.}~\bibnamefont {Lämmerzahl}}, \bibinfo {author}
  {\bibfnamefont {K.}~\bibnamefont {Sengstock}}, \bibinfo {author}
  {\bibfnamefont {N.}~\bibnamefont {Gaaloul}}, \bibinfo {author} {\bibfnamefont
  {W.}~\bibnamefont {Ertmer}},\ and\ \bibinfo {author} {\bibfnamefont {E.~M.}\
  \bibnamefont {Rasel}},\ }\href
  {https://doi.org/10.1088/1367-2630/17/6/065001} {\bibfield  {journal}
  {\bibinfo  {journal} {New Journal of Physics}\ }\textbf {\bibinfo {volume}
  {17}},\ \bibinfo {pages} {065001} (\bibinfo {year} {2015})}\BibitemShut
  {NoStop}%
\bibitem [{\citenamefont {Heine}\ \emph {et~al.}(2020)\citenamefont {Heine},
  \citenamefont {Matthias}, \citenamefont {Sahelgozin}, \citenamefont {Herr},
  \citenamefont {Abend}, \citenamefont {Timmen}, \citenamefont {Müller},\ and\
  \citenamefont {Rasel}}]{heine_transportable_2020}%
  \BibitemOpen
  \bibfield  {author} {\bibinfo {author} {\bibfnamefont {N.}~\bibnamefont
  {Heine}}, \bibinfo {author} {\bibfnamefont {J.}~\bibnamefont {Matthias}},
  \bibinfo {author} {\bibfnamefont {M.}~\bibnamefont {Sahelgozin}}, \bibinfo
  {author} {\bibfnamefont {W.}~\bibnamefont {Herr}}, \bibinfo {author}
  {\bibfnamefont {S.}~\bibnamefont {Abend}}, \bibinfo {author} {\bibfnamefont
  {L.}~\bibnamefont {Timmen}}, \bibinfo {author} {\bibfnamefont
  {J.}~\bibnamefont {Müller}},\ and\ \bibinfo {author} {\bibfnamefont {E.~M.}\
  \bibnamefont {Rasel}},\ }\href {https://doi.org/10.1140/epjd/e2020-10120-x}
  {\bibfield  {journal} {\bibinfo  {journal} {The European Physical Journal D}\
  }\textbf {\bibinfo {volume} {74}},\ \bibinfo {pages} {174} (\bibinfo {year}
  {2020})}\BibitemShut {NoStop}%
\bibitem [{\citenamefont {Deppner}\ \emph {et~al.}(2021)\citenamefont
  {Deppner}, \citenamefont {Herr}, \citenamefont {Cornelius}, \citenamefont
  {Stromberger}, \citenamefont {Sternke}, \citenamefont {Grzeschik},
  \citenamefont {Grote}, \citenamefont {Rudolph}, \citenamefont {Herrmann},
  \citenamefont {Krutzik}, \citenamefont {Wenzlawski}, \citenamefont {Corgier},
  \citenamefont {Charron}, \citenamefont {Gu{\'e}ry-Odelin}, \citenamefont
  {Gaaloul}, \citenamefont {Lämmerzahl}, \citenamefont {Peters}, \citenamefont
  {Windpassinger},\ and\ \citenamefont {Rasel}}]{deppner_collective-mode_2021}%
  \BibitemOpen
  \bibfield  {author} {\bibinfo {author} {\bibfnamefont {C.}~\bibnamefont
  {Deppner}}, \bibinfo {author} {\bibfnamefont {W.}~\bibnamefont {Herr}},
  \bibinfo {author} {\bibfnamefont {M.}~\bibnamefont {Cornelius}}, \bibinfo
  {author} {\bibfnamefont {P.}~\bibnamefont {Stromberger}}, \bibinfo {author}
  {\bibfnamefont {T.}~\bibnamefont {Sternke}}, \bibinfo {author} {\bibfnamefont
  {C.}~\bibnamefont {Grzeschik}}, \bibinfo {author} {\bibfnamefont
  {A.}~\bibnamefont {Grote}}, \bibinfo {author} {\bibfnamefont
  {J.}~\bibnamefont {Rudolph}}, \bibinfo {author} {\bibfnamefont
  {S.}~\bibnamefont {Herrmann}}, \bibinfo {author} {\bibfnamefont
  {M.}~\bibnamefont {Krutzik}}, \bibinfo {author} {\bibfnamefont
  {A.}~\bibnamefont {Wenzlawski}}, \bibinfo {author} {\bibfnamefont
  {R.}~\bibnamefont {Corgier}}, \bibinfo {author} {\bibfnamefont
  {E.}~\bibnamefont {Charron}}, \bibinfo {author} {\bibfnamefont
  {D.}~\bibnamefont {Gu{\'e}ry-Odelin}}, \bibinfo {author} {\bibfnamefont
  {N.}~\bibnamefont {Gaaloul}}, \bibinfo {author} {\bibfnamefont
  {C.}~\bibnamefont {Lämmerzahl}}, \bibinfo {author} {\bibfnamefont
  {A.}~\bibnamefont {Peters}}, \bibinfo {author} {\bibfnamefont
  {P.}~\bibnamefont {Windpassinger}},\ and\ \bibinfo {author} {\bibfnamefont
  {E.~M.}\ \bibnamefont {Rasel}},\ }\href
  {https://doi.org/10.1103/PhysRevLett.127.100401} {\bibfield  {journal}
  {\bibinfo  {journal} {Physical Review Letters}\ }\textbf {\bibinfo {volume}
  {127}},\ \bibinfo {pages} {100401} (\bibinfo {year} {2021})}\BibitemShut
  {NoStop}%
\bibitem [{\citenamefont {Bjorklund}\ \emph {et~al.}(1983)\citenamefont
  {Bjorklund}, \citenamefont {Levenson}, \citenamefont {Lenth},\ and\
  \citenamefont {Ortiz}}]{bjorklund_frequency_1983}%
  \BibitemOpen
  \bibfield  {author} {\bibinfo {author} {\bibfnamefont {G.~C.}\ \bibnamefont
  {Bjorklund}}, \bibinfo {author} {\bibfnamefont {M.~D.}\ \bibnamefont
  {Levenson}}, \bibinfo {author} {\bibfnamefont {W.}~\bibnamefont {Lenth}},\
  and\ \bibinfo {author} {\bibfnamefont {C.}~\bibnamefont {Ortiz}},\ }\href
  {https://doi.org/10.1007/BF00688820} {\bibfield  {journal} {\bibinfo
  {journal} {Applied Physics B}\ }\textbf {\bibinfo {volume} {32}},\ \bibinfo
  {pages} {145} (\bibinfo {year} {1983})}\BibitemShut {NoStop}%
\bibitem [{\citenamefont {Saywell}\ \emph {et~al.}(2023)\citenamefont
  {Saywell}, \citenamefont {Carey}, \citenamefont {Light}, \citenamefont
  {Szigeti}, \citenamefont {Milne}, \citenamefont {Gill}, \citenamefont {Goh},
  \citenamefont {Perunicic}, \citenamefont {Wilson}, \citenamefont {Macrae},
  \citenamefont {Rischka}, \citenamefont {Everitt}, \citenamefont {Robins},
  \citenamefont {Anderson}, \citenamefont {Hush},\ and\ \citenamefont
  {Biercuk}}]{saywell_enhancing_2023}%
  \BibitemOpen
  \bibfield  {author} {\bibinfo {author} {\bibfnamefont {J.~C.}\ \bibnamefont
  {Saywell}}, \bibinfo {author} {\bibfnamefont {M.~S.}\ \bibnamefont {Carey}},
  \bibinfo {author} {\bibfnamefont {P.~S.}\ \bibnamefont {Light}}, \bibinfo
  {author} {\bibfnamefont {S.~S.}\ \bibnamefont {Szigeti}}, \bibinfo {author}
  {\bibfnamefont {A.~R.}\ \bibnamefont {Milne}}, \bibinfo {author}
  {\bibfnamefont {K.~S.}\ \bibnamefont {Gill}}, \bibinfo {author}
  {\bibfnamefont {M.~L.}\ \bibnamefont {Goh}}, \bibinfo {author} {\bibfnamefont
  {V.~S.}\ \bibnamefont {Perunicic}}, \bibinfo {author} {\bibfnamefont {N.~M.}\
  \bibnamefont {Wilson}}, \bibinfo {author} {\bibfnamefont {C.~D.}\
  \bibnamefont {Macrae}}, \bibinfo {author} {\bibfnamefont {A.}~\bibnamefont
  {Rischka}}, \bibinfo {author} {\bibfnamefont {P.~J.}\ \bibnamefont
  {Everitt}}, \bibinfo {author} {\bibfnamefont {N.~P.}\ \bibnamefont {Robins}},
  \bibinfo {author} {\bibfnamefont {R.~P.}\ \bibnamefont {Anderson}}, \bibinfo
  {author} {\bibfnamefont {M.~R.}\ \bibnamefont {Hush}},\ and\ \bibinfo
  {author} {\bibfnamefont {M.~J.}\ \bibnamefont {Biercuk}},\ }\href
  {https://doi.org/10.1038/s41467-023-43374-0} {\bibfield  {journal} {\bibinfo
  {journal} {Nature Communications}\ }\textbf {\bibinfo {volume} {14}},\
  \bibinfo {pages} {7626} (\bibinfo {year} {2023})},\ \bibinfo {note} {number:
  1 Publisher: Nature Publishing Group}\BibitemShut {NoStop}%
\bibitem [{Note1()}]{Note1}%
  \BibitemOpen
  \bibinfo {note} {For $^{87}$Rb atoms starting in the $\mathinner
  {|{F=2,m_F=0}\rangle }$ state, the timescale for optical pumping into the
  $\mathinner {|{2,2}\rangle }$ state is $\mathord {\sim }\SI {10}{\micro
  \second }$ for a resonant laser whose intensity is one-tenth the saturation
  intensity.}\BibitemShut {Stop}%
\bibitem [{\citenamefont {Metcalf}\ and\ \citenamefont {Van~der
  Straten}()}]{laser_cooling_and_trapping}%
  \BibitemOpen
  \bibfield  {author} {\bibinfo {author} {\bibfnamefont {H.~J.}\ \bibnamefont
  {Metcalf}}\ and\ \bibinfo {author} {\bibfnamefont {P.}~\bibnamefont {Van~der
  Straten}},\ }\href@noop {} {\emph {\bibinfo {title} {Laser {Cooling} and
  {Trapping}}}}\BibitemShut {NoStop}%
\bibitem [{\citenamefont {McIntyre}(1966)}]{mcintyre_multiplication_1966}%
  \BibitemOpen
  \bibfield  {author} {\bibinfo {author} {\bibfnamefont {R.}~\bibnamefont
  {McIntyre}},\ }\href {https://doi.org/10.1109/T-ED.1966.15651} {\bibfield
  {journal} {\bibinfo  {journal} {IEEE Transactions on Electron Devices}\
  }\textbf {\bibinfo {volume} {ED-13}},\ \bibinfo {pages} {164} (\bibinfo
  {year} {1966})}\BibitemShut {NoStop}%
\bibitem [{\citenamefont {Li}\ \emph {et~al.}(2023)\citenamefont {Li},
  \citenamefont {Long}, \citenamefont {Huang}, \citenamefont {Chen},
  \citenamefont {Yang}, \citenamefont {Jiang}, \citenamefont {Xiang},
  \citenamefont {Ma}, \citenamefont {He}, \citenamefont {Chen},\ and\
  \citenamefont {Chen}}]{li_continuous_2023}%
  \BibitemOpen
  \bibfield  {author} {\bibinfo {author} {\bibfnamefont {C.-Y.}\ \bibnamefont
  {Li}}, \bibinfo {author} {\bibfnamefont {J.-B.}\ \bibnamefont {Long}},
  \bibinfo {author} {\bibfnamefont {M.-Q.}\ \bibnamefont {Huang}}, \bibinfo
  {author} {\bibfnamefont {B.}~\bibnamefont {Chen}}, \bibinfo {author}
  {\bibfnamefont {Y.-M.}\ \bibnamefont {Yang}}, \bibinfo {author}
  {\bibfnamefont {X.}~\bibnamefont {Jiang}}, \bibinfo {author} {\bibfnamefont
  {C.-F.}\ \bibnamefont {Xiang}}, \bibinfo {author} {\bibfnamefont {Z.-L.}\
  \bibnamefont {Ma}}, \bibinfo {author} {\bibfnamefont {D.-Q.}\ \bibnamefont
  {He}}, \bibinfo {author} {\bibfnamefont {L.-K.}\ \bibnamefont {Chen}},\ and\
  \bibinfo {author} {\bibfnamefont {S.}~\bibnamefont {Chen}},\ }\href
  {https://doi.org/10.1103/PhysRevA.108.032811} {\bibfield  {journal} {\bibinfo
   {journal} {Physical Review A}\ }\textbf {\bibinfo {volume} {108}},\ \bibinfo
  {pages} {032811} (\bibinfo {year} {2023})}\BibitemShut {NoStop}%
\bibitem [{\citenamefont {Hardman}\ \emph
  {et~al.}(2016{\natexlab{b}})\citenamefont {Hardman}, \citenamefont {Everitt},
  \citenamefont {McDonald}, \citenamefont {Manju}, \citenamefont {Wigley},
  \citenamefont {Sooriyabandara}, \citenamefont {Kuhn}, \citenamefont {Debs},
  \citenamefont {Close},\ and\ \citenamefont
  {Robins}}]{hardman_simultaneous_2016}%
  \BibitemOpen
  \bibfield  {author} {\bibinfo {author} {\bibfnamefont {K.~S.}\ \bibnamefont
  {Hardman}}, \bibinfo {author} {\bibfnamefont {P.~J.}\ \bibnamefont
  {Everitt}}, \bibinfo {author} {\bibfnamefont {G.~D.}\ \bibnamefont
  {McDonald}}, \bibinfo {author} {\bibfnamefont {P.}~\bibnamefont {Manju}},
  \bibinfo {author} {\bibfnamefont {P.~B.}\ \bibnamefont {Wigley}}, \bibinfo
  {author} {\bibfnamefont {M.~A.}\ \bibnamefont {Sooriyabandara}}, \bibinfo
  {author} {\bibfnamefont {C.~C.~N.}\ \bibnamefont {Kuhn}}, \bibinfo {author}
  {\bibfnamefont {J.~E.}\ \bibnamefont {Debs}}, \bibinfo {author}
  {\bibfnamefont {J.~D.}\ \bibnamefont {Close}},\ and\ \bibinfo {author}
  {\bibfnamefont {N.~P.}\ \bibnamefont {Robins}},\ }\href
  {https://doi.org/10.1103/PhysRevLett.117.138501} {\bibfield  {journal}
  {\bibinfo  {journal} {Physical Review Letters}\ }\textbf {\bibinfo {volume}
  {117}},\ \bibinfo {pages} {138501} (\bibinfo {year}
  {2016}{\natexlab{b}})}\BibitemShut {NoStop}%
\bibitem [{\citenamefont {Wigley}\ \emph {et~al.}(2019)\citenamefont {Wigley},
  \citenamefont {Hardman}, \citenamefont {Freier}, \citenamefont {Everitt},
  \citenamefont {Legge}, \citenamefont {Manju}, \citenamefont {Close},\ and\
  \citenamefont {Robins}}]{wigley_readout-delay-free_2019}%
  \BibitemOpen
  \bibfield  {author} {\bibinfo {author} {\bibfnamefont {P.~B.}\ \bibnamefont
  {Wigley}}, \bibinfo {author} {\bibfnamefont {K.~S.}\ \bibnamefont {Hardman}},
  \bibinfo {author} {\bibfnamefont {C.}~\bibnamefont {Freier}}, \bibinfo
  {author} {\bibfnamefont {P.~J.}\ \bibnamefont {Everitt}}, \bibinfo {author}
  {\bibfnamefont {S.}~\bibnamefont {Legge}}, \bibinfo {author} {\bibfnamefont
  {P.}~\bibnamefont {Manju}}, \bibinfo {author} {\bibfnamefont {J.~D.}\
  \bibnamefont {Close}},\ and\ \bibinfo {author} {\bibfnamefont {N.~P.}\
  \bibnamefont {Robins}},\ }\href {https://doi.org/10.1103/PhysRevA.99.023615}
  {\bibfield  {journal} {\bibinfo  {journal} {Physical Review A}\ }\textbf
  {\bibinfo {volume} {99}},\ \bibinfo {pages} {023615} (\bibinfo {year}
  {2019})}\BibitemShut {NoStop}%
\bibitem [{\citenamefont {Wang}\ \emph {et~al.}(2023)\citenamefont {Wang},
  \citenamefont {Tong}, \citenamefont {Xie}, \citenamefont {Wang},
  \citenamefont {Feng},\ and\ \citenamefont {Wang}}]{wang_enhanced_2023}%
  \BibitemOpen
  \bibfield  {author} {\bibinfo {author} {\bibfnamefont {J.}~\bibnamefont
  {Wang}}, \bibinfo {author} {\bibfnamefont {J.}~\bibnamefont {Tong}}, \bibinfo
  {author} {\bibfnamefont {W.}~\bibnamefont {Xie}}, \bibinfo {author}
  {\bibfnamefont {Z.}~\bibnamefont {Wang}}, \bibinfo {author} {\bibfnamefont
  {Y.}~\bibnamefont {Feng}},\ and\ \bibinfo {author} {\bibfnamefont
  {X.}~\bibnamefont {Wang}},\ }\href {https://doi.org/10.3390/s23115071}
  {\bibfield  {journal} {\bibinfo  {journal} {Sensors}\ }\textbf {\bibinfo
  {volume} {23}},\ \bibinfo {pages} {5071} (\bibinfo {year} {2023})},\ \bibinfo
  {note} {number: 11 Publisher: Multidisciplinary Digital Publishing
  Institute}\BibitemShut {NoStop}%
\bibitem [{\citenamefont {Ben-Aïcha}\ \emph {et~al.}(2024)\citenamefont
  {Ben-Aïcha}, \citenamefont {Mehdi}, \citenamefont {Freier}, \citenamefont
  {Szigeti}, \citenamefont {Wigley}, \citenamefont {Conlon}, \citenamefont
  {Husband}, \citenamefont {Legge}, \citenamefont {Eagle}, \citenamefont
  {Hope}, \citenamefont {Robins}, \citenamefont {Close}, \citenamefont
  {Hardman}, \citenamefont {Haine},\ and\ \citenamefont
  {Thomas}}]{ben-aicha_dual_2024}%
  \BibitemOpen
  \bibfield  {author} {\bibinfo {author} {\bibfnamefont {Y.}~\bibnamefont
  {Ben-Aïcha}}, \bibinfo {author} {\bibfnamefont {Z.}~\bibnamefont {Mehdi}},
  \bibinfo {author} {\bibfnamefont {C.}~\bibnamefont {Freier}}, \bibinfo
  {author} {\bibfnamefont {S.~S.}\ \bibnamefont {Szigeti}}, \bibinfo {author}
  {\bibfnamefont {P.~B.}\ \bibnamefont {Wigley}}, \bibinfo {author}
  {\bibfnamefont {L.~O.}\ \bibnamefont {Conlon}}, \bibinfo {author}
  {\bibfnamefont {R.}~\bibnamefont {Husband}}, \bibinfo {author} {\bibfnamefont
  {S.}~\bibnamefont {Legge}}, \bibinfo {author} {\bibfnamefont {R.~H.}\
  \bibnamefont {Eagle}}, \bibinfo {author} {\bibfnamefont {J.~J.}\ \bibnamefont
  {Hope}}, \bibinfo {author} {\bibfnamefont {N.~P.}\ \bibnamefont {Robins}},
  \bibinfo {author} {\bibfnamefont {J.~D.}\ \bibnamefont {Close}}, \bibinfo
  {author} {\bibfnamefont {K.~S.}\ \bibnamefont {Hardman}}, \bibinfo {author}
  {\bibfnamefont {S.~A.}\ \bibnamefont {Haine}},\ and\ \bibinfo {author}
  {\bibfnamefont {R.~J.}\ \bibnamefont {Thomas}},\ }\href
  {https://doi.org/10.48550/arXiv.2405.00400} {\bibinfo {title} {A {Dual}
  {Open} {Atom} {Interferometer} for {Compact}, {Mobile} {Quantum} {Sensing}}}
  (\bibinfo {year} {2024}),\ \bibinfo {note} {arXiv:2405.00400 [physics,
  physics:quant-ph]}\BibitemShut {NoStop}%
\bibitem [{\citenamefont {Szigeti}\ \emph {et~al.}(2021)\citenamefont
  {Szigeti}, \citenamefont {Hosten},\ and\ \citenamefont
  {Haine}}]{szigeti_improving_2021}%
  \BibitemOpen
  \bibfield  {author} {\bibinfo {author} {\bibfnamefont {S.~S.}\ \bibnamefont
  {Szigeti}}, \bibinfo {author} {\bibfnamefont {O.}~\bibnamefont {Hosten}},\
  and\ \bibinfo {author} {\bibfnamefont {S.~A.}\ \bibnamefont {Haine}},\ }\href
  {https://doi.org/10.1063/5.0050235} {\bibfield  {journal} {\bibinfo
  {journal} {Applied Physics Letters}\ }\textbf {\bibinfo {volume} {118}},\
  \bibinfo {pages} {140501} (\bibinfo {year} {2021})}\BibitemShut {NoStop}%
\bibitem [{\citenamefont {Wineland}\ \emph {et~al.}(1994)\citenamefont
  {Wineland}, \citenamefont {Bollinger}, \citenamefont {Itano},\ and\
  \citenamefont {Heinzen}}]{wineland_squeezed_1994}%
  \BibitemOpen
  \bibfield  {author} {\bibinfo {author} {\bibfnamefont {D.~J.}\ \bibnamefont
  {Wineland}}, \bibinfo {author} {\bibfnamefont {J.~J.}\ \bibnamefont
  {Bollinger}}, \bibinfo {author} {\bibfnamefont {W.~M.}\ \bibnamefont
  {Itano}},\ and\ \bibinfo {author} {\bibfnamefont {D.~J.}\ \bibnamefont
  {Heinzen}},\ }\href {https://doi.org/10.1103/PhysRevA.50.67} {\bibfield
  {journal} {\bibinfo  {journal} {Phys. Rev. A}\ }\textbf {\bibinfo {volume}
  {50}},\ \bibinfo {pages} {67} (\bibinfo {year} {1994})}\BibitemShut {NoStop}%
\end{thebibliography}

\end{document}